\documentclass[12pt]{article}
\pdfoutput=1
\usepackage{latexsym,cite}
\usepackage{amsmath}
\usepackage{amssymb}
\usepackage{amsthm}
\usepackage{subcaption}
\usepackage{mathtools}
\usepackage{braket}

\numberwithin{equation}{section}

\usepackage[top = 1. in,bottom = 0.93 in, left = 1 in, right=0.93 in]{geometry}

\newcommand{\arXiv}[1]{\href{http://www.arXiv.org/abs/#1}{arXiv:#1}}
\newcommand{\doi}[2]{\href{http://dx.doi.org/#2}{#1}}
\usepackage[colorlinks=true, linkcolor=blue, bookmarks=true]{hyperref}

\allowdisplaybreaks

\makeatletter
\renewcommand\section{\@startsection {section}{1}{\z@}%
                  {-3.5ex \@plus -1ex \@minus -.2ex}
                  {2.3ex \@plus.2ex}%
                  {\normalfont\large\bfseries}}
\renewcommand\subsection{\@startsection{subsection}{2}{\z@}%
                   {-3.25ex\@plus -1ex \@minus -.2ex}%
                   {1.5ex \@plus .2ex}%
                   {\normalfont\bfseries}}
\makeatother

\newcommand{\beq}{\begin{equation}}
\newcommand{\eeq}{\end{equation}}

\newcommand{\de}{\delta}

\newcommand{\ena}{\end{eqnarray}}
\newcommand{\beqa}{\begin{eqnarray}}
\newcommand{\eeqa}{\end{eqnarray}}
\newcommand{\bea}{\begin{eqnarray}}
\newcommand{\eea}{\end{eqnarray}}
\newcommand{\al}{\alpha}
\newcommand{\ab}{\bar\alpha}
\newcommand{\ad}{a^\dagger}
\newcommand{\Sz}{Szeg\H{o}}
\newcommand{\Tr}{\mathrm{Tr}}
\newcommand{\nn}{\nonumber}
\newcommand{\Hm}{H_{\min}}

\begin{document}
\begin{titlepage}
\begin{flushright}
\phantom{arXiv:yymm.nnnn}
\end{flushright}
\vspace{15mm}
\begin{center}
{\huge\bf A~superintegrable~quantum~field~theory}
\vskip 10mm
{\large Marine De Clerck}
\vskip 3mm
{\it DAMTP, University of Cambridge, Cambridge, United Kingdom}
\vskip 7mm
{\large Oleg Evnin}
\vskip 3mm
{\it  High Energy Physics Research Unit, Faculty of Science, \\Chulalongkorn University,
Bangkok, Thailand\vspace{1mm} \\ Theoretische Natuurkunde, Vrije Universiteit Brussel (VUB) \\
\&\,\,
International Solvay Institutes, Brussels, Belgium}
\vskip 7mm
{\small {\tt md989@cam.ac.uk, oleg.evnin@gmail.com}}
\vskip 35mm
{\bf ABSTRACT}\vspace{3mm}
\end{center}
G\'erard and Grellier proposed, under the name of the cubic \Sz{} equation, a remarkable classical field theory on a circle with a quartic Hamiltonian.
The Lax integrability structure that emerges from their definition is so constraining that it allows for writing down an explicit general
solution for prescribed initial data, and at the same time, the dynamics is highly nontrivial and involves turbulent energy transfer
to arbitrarily short wavelengths. The quantum version of the same Hamiltonian is even more striking: not only the Hamiltonian itself,
but also its associated conserved hierarchies display purely integer spectra, indicating a structure beyond ordinary quantum integrability.
Here, we initiate a systematic study of this quantum system by presenting a mixture of analytic results and empirical observations on
the structure of its eigenvalues and eigenvectors, conservation laws, ladder operators, etc.

\vfill

\end{titlepage}


\section{Introduction}

Our main concern here will be with the quantum Hamiltonian
\beq\label{HGG}
H\equiv \frac12\sum_{n+m=k+l}^{n,m,k,l\ge 0} a^\dagger_n a^\dagger_m a_k a_l,
\eeq
where $a^\dagger_n$ and $a_n$ with $n\ge 0$ are a set of usual bosonic creation-annihilation operators satisfying $[a_n, a_m^\dagger]=\delta_{mn}$.
The structure is that of a typical particle-number-conserving (nonrelativistic) quantum field theory with quartic interactions, as one may commonly see in theoretical condensed matter physics, yet the couplings
between the modes are chosen in such a way as to impart the system a rich and elaborate integrability structure. Our goal is to investigate this structure.

The classical version of (\ref{HGG}) is the Hamiltonian
\beq\label{Hcl}
H_{cl}\equiv \frac12\sum_{n+m=k+l}^{n,m,k,l\ge 0} \bar\al_n \bar\al_m \al_k \al_l
\eeq
for complex canonical coordinates $\al_n(t)$ whose conjugate momenta are $i\bar\al_n(t)$, with bars denoting the ordinary complex conjugation.
The Hamiltonian (\ref{Hcl}) was introduced, studied and solved in a series of works by Patrick G\'erard and Sandrine Grellier \cite{GG,GG1,GG2,GG3}, using position space notation that we shall explain below (as opposed to our mode space notation). We shall be referring to (\ref{HGG}) and (\ref{Hcl}) as G\'erard-Grellier (GG) Hamiltonians.

The main motivation for introducing and studying (\ref{Hcl}) was the topic of turbulent energy transfer from modes with long wavelength (small $n$) to those with short wavelength (large $n$). Such turbulent phenomena are an important topic in contemporary PDE mathematics, and are typically challenging to analyze -- see \cite{nls} for a well-known example where a ``weak weak'' form of turbulence is established rigorously with considerable effort. The GG Hamiltonian (\ref{Hcl}) plays a very special role in this regard since it displays an array of turbulent phenomena that can be explicitly analyzed using integrability methods. We shall review the classical integrability of (\ref{Hcl}) in section~\ref{seccl} after spelling out in section~\ref{secres} its place in the broader context of resonant Hamiltonian systems with their connections to weakly nonlinear equations appearing in physics.

While in-depth understanding is available for the integrable structure of the classical Hamiltonian (\ref{Hcl}) with its Lax pairs and hierarchies of conservation laws,
this knowledge does not directly transfer to the quantum system, as naive quantization of the classical expressions without special attention to the ordering ambiguities does
not recover the conserved charges of the quantum system. Perhaps the most striking feature of the quantum Hamiltonian (\ref{HGG}), first pointed out in \cite{quantres}, is that its eigenvalue spectrum is made of integers.
This feature is beyond any regular expectations, even for a quantum integrable system. Indeed, while a fundamental definition of quantum integrability is lacking \cite{quantint1,quantint2}, the general expectation is that the energy levels of a typical quantum integrable system look like independent random numbers thrown on the real line \cite{DKPR}, without level correlations or level repulsion. The perfect integers with rigid level spacings are clearly a different story, and such spectra are typically seen in superintegrable quantum-mechanical systems \cite{mapping}. In view of this picture, we would like to refer to (\ref{HGG}) as a superintegrable quantum field theory.

There is more to the story. In what follows, we shall present some approaches to constructing quantum conservation laws. One example is
\beq
	H_{\min} = \hspace{-2mm}\sum_{\substack{n,m,k,l = 1 \\ n+m=k+l}}^\infty\hspace{-2mm} \min(n,m,k,l) \ a^\dagger_n a^\dagger_{m} a_k a_{l} +  \sum_{k=1}^\infty k^2 a^\dagger_k a_k,
	\label{Hmin}
\eeq
first pointed out in \cite{bound}. This operator commutes with (\ref{HGG}), and there will be further operators of this sort as well. All the conservation laws we find
have purely integer spectra.

In what follows, we shall summarize our current understanding of the quantum G\'erard-Grellier Hamiltonian, its hierarchy of conservation laws, and their spectra.
This investigation will combine analytic and numerical tools.


\section{The context of resonant Hamiltonian systems}\label{secres}

Before focusing on the GG Hamiltonian (\ref{HGG}), it could be good to step back for a moment and examine the more general system
\beq\label{Hres}
H_{res}\equiv \frac12\sum_{n+m=k+l}^{n,m,k,l\ge 0}C_{nmkl} \bar\al_n \bar\al_m \al_k \al_l
\eeq
with arbitrary coupling coefficients $C$ satisfying $C_{nmkl}=C_{nmlk}=\bar C_{klnm}$. Evidently, setting $C$ to 1 leads to (\ref{Hcl}), while subsequent canonical quantization leads to (\ref{HGG}). We refer to such Hamiltonians as {\it resonant Hamiltonians} because of the presence of the resonance condition $n+m=k+l$.

Hamiltonian systems of the form (\ref{Hres}) emerge naturally from weakly nonlinear approximations to Hamiltonian PDEs whose linearized normal modes possess
a perfectly resonant, equispaced spectrum of frequencies \cite{resrev}. Examples include studies of nonlinear Schr\"odinger equation with harmonic potentials \cite{GHT,GT,BBCE,GGT,fennell,BEF,Schwinte}, which is a model for trapped Bose-Einstein condensates \cite{fetter}, as well as studies of waves in asymptotically anti-de Sitter spacetimes \cite{FPU,CEV1,CEV2,BMR,CF,BEL}, with and without gravitational interactions, in relation to the anti-de Sitter instability conjecture of Bizo\'n and Rostworowski \cite{BR}. These systems have also been studied in their own right, mostly as a dynamical arena for turbulent transfer of excitations from $\alpha_n$ with small $n$ to high $n$ (long wavelength to short wavelength in physical settings). This line of research includes the original articles \cite{GG,GG1,GG2,GG3} that introduced and studied (\ref{Hcl}), as well as \cite{Xu,solvable,breathing,cascade,Biasi,BG}. Because of the presence
of two conservation laws
\beq\label{NMcl}
N=\sum_{n=0}^\infty |\alpha_n|^2,\qquad M=\sum_{n=1}^\infty n |\alpha_n|^2,
\eeq
transfer of excitations to higher $n$ is necessarily accompanied by transfer of excitations to low $n$, which is known as a `dual cascade' \cite{dual}, see \cite{Biasi,BG} for analytically tractable examples.

For a quick demonstration of how a Hamiltonian system of the form (\ref{Hres}) can emerge from a realistic, physical PDE, consider the one-dimensional nonlinear Schr\"odinger equation with a harmonic potential:
\begin{equation}
i\,\frac{\partial \Psi}{\partial t}=\frac12\left(-\frac{\partial^2}{\partial x^2}+x^2\right)\Psi +g|\Psi|^2\Psi.
\label{NLS1d}
\end{equation}
With the nonlinearity turned off ($g=0$), the general solution is
\begin{equation}
\Psi=\sum_{n=0}^\infty \alpha_n \psi_n(x) e^{-iE_n t},\qquad E_n=n+\frac12,\qquad \frac12\left(-\frac{\partial^2}{\partial x^2}+x^2\right)\psi_n=E_n\psi_n,
\label{NLS1dlin}
\end{equation}
with constants $\alpha_n$. At small nonzero coupling $g$,  $\alpha_n$ are no longer constant in time and acquire slow variations. Assuming thus that $\alpha_n$ are functions of time, we can substitute \eqref{NLS1dlin} into \eqref{NLS1d} to obtain
\begin{equation} 
 i \,\frac{d\alpha_n}{dt}  = g\sum_{k,l,m=0}^\infty C_{nmkl} \,\bar \alpha_m \alpha_k  \alpha_l \,e^{i(E_n+E_m-E_k-E_l)t}, 
\label{ampltd}
\end{equation}
where $C_{nmkl}=\int dx \,\psi_n  \psi_m \psi_k \psi_l$. At small $g$, $\alpha_n$ vary slowly, while the last exponential factor oscillates on time scales of order 1.
It is legitimate to expect that these `fast' oscillations average out and only resonant terms with $E_n+E_m-E_k-E_l=n+m-k-l=0$ contribute substantially to the evolution.
Technically, this provides a good approximation up to $t\sim 1/g$ (but not on longer timescales). The result of eliminating nonresonant terms is precisely the Hamiltonian equation of motion of (\ref{Hres}), after $g$ has been absorbed into a redefinition of time. Related mathematically rigorous proofs protecting the accuracy of this {\it resonant approximation} can be found in \cite{GHT, fennell}. If one wants to obtain the GG Hamiltonian (\ref{Hcl}) via this resonant approximation process, one has to start with the so-called half-wave equation on a circle \cite{halfwave}.

Only special values of the couplings $C$ arise from common physical PDEs, so an interesting question is how to engineer systems that produce desired values of couplings (including the ones we shall focus on in this paper). Ultracold atomic gases in harmonic traps provide a good setting, since harmonic potentials give rise to resonant structures
necessary for (\ref{Hres}). An outstanding problem is, however, to tune the couplings between the normal modes (for instance, by incorporating position-dependent nonlinearities). Another option is waves on a circle, but those waves will have to move in one direction, since the wavenumber index $n$ in $\alpha_n$ is nonnegative. Such `chiral waves' are known to exist as edge states in some condensed matter system \cite{edge}. Once again, the interactions of individual normal modes will have to be differentially controlled.

Canonical quantization of (\ref{Hres}) leads to the corresponding quantum Hamiltonian
\beq\label{Hresq}
H_{quant}= \frac12\sum_{n+m=k+l}^{n,m,k,l\ge 0}C_{nmkl} a^\dagger_n a^\dagger_m a_k a_l,
\eeq
with $[a_n, a_m^\dagger]=\delta_{mn}$.
This Hamiltonian is strikingly simple to work with, at least at the level of numerical diagonalization, independent of the actual values of the couplings $C$. Indeed,
quantization of (\ref{NMcl}) immediately leads to two quantum conservation laws
\beq\label{NMq}
N=\sum_{n=0}^\infty a^\dagger_n a_n,\qquad M=\sum_{n=1}^\infty n  \,a^\dagger_n a_n,
\eeq
while the space of states is spanned by the Fock vectors: there is one such vector for each set of occupation numbers $\eta_n$, and it is obtained by acting on the vacuum $|0\rangle$ (annihilated by all $a_n$) with the creation operators $a^\dagger_n$ as $(a_0^\dagger)^{\eta_0}(a_1^\dagger)^{\eta_1}\cdots|0\rangle$. But there is only a finite number of choices of $\eta_n$ for each $N$ and $M$ (given by the number of integer partitions of $M$ into at most $N$ parts), while all matrix elements of $H_{quant}$ between Fock states with different $N$ or $M$ must vanish. For that reason, diagonalization of (\ref{Hresq}) is reduced to diagonalizing an infinite family of finite-sized matrices, which makes the eigenvalues directly accessible \cite{quantres}. This is in fact how the integer eigenvalues of (\ref{HGG}) were originally observed. Note that all of this has nothing to do with solvability of (\ref{Hres}) and (\ref{Hresq}) and the structure is present for any values of the couplings, despite the fact that the corresponding classical dynamics may be arbitrarily complicated (chaotic, turbulent, etc). For special values of $C$ corresponding to quantum integrable systems, as in (\ref{HGG}), there is much more structure still.

Just like its classical counterpart (\ref{Hres}), the quantum Hamiltonian (\ref{Hresq}) arises as a weak coupling approximation to certain quantum field theories,
and this is even more intuitive than the classical story \cite{HEPMAD, shift, qperiod1, qperiod2}. Indeed, when the classical normal modes have commensurate frequencies, the corresponding quanta have commensurate energies. As a result, quantization of free field theories with highly resonant spectra (nonrelativistic Schr\"odinger fields in harmonic traps \cite{shift,qperiod1}, relativistic fields in anti-de Sitter spacetimes \cite{HEPMAD,qperiod2}, etc) leads to highly degenerate multiparticle energy spectra.
When weak interactions are turned on, these degenerate levels split, and $H_{quant}$ is the Hamiltonian one needs to diagonalize, in line with the standard Rayleigh–Schr\"odinger perturbation theory for degenerate spectra, to obtain the energy shifts that split the degenerate noninteracting energy levels. This picture has led to studies of the so-called lowest Landau level Hamiltonian, which belongs to the class (\ref{Hresq}) in relation to the physics of trapped ultracold atomic gases \cite{qLLL1,qLLL2,qLLL3,qLLL4,qLLLexp}.


\section{Classical integrability of the G\'erard-Grellier Hamiltonian}\label{seccl}

Before proceeding to the quantum case (\ref{HGG}), it would be wise to review the classical integrability theory of (\ref{Hcl}) developed in detail in the original
works \cite{GG,GG1,GG2,GG3}. In doing so, we will also attempt to adapt the presentation for the eyes of physicists.

The equations of motion corresponding to (\ref{Hcl}) read
\beq\label{SzEOM}
i\dot\alpha_n=\sum_{m=0}^\infty\sum_{k=0}^{n+m} \bar\alpha_m\al_k\al_{n+m-k},
\eeq
where dots will denote time derivatives.
For some purposes, it is convenient to rewrite these equations in terms of the position space field $u(\theta)$ on the unit circle $\theta\in[0,2\pi)$:
\beq\label{udef}
u(\theta,t)\equiv \sum_{n=0}^\infty e^{in\theta}\alpha_n(t),
\eeq
which yields
\beq\label{uSz}
i\dot u=\Pi(|u|^2u),
\eeq
with 
\beq
\Pi\left(\sum_{n=-\infty}^\infty h_n e^{in\theta}\right)\equiv\sum_{n=0}^\infty h_n e^{in\theta}
\eeq
 being the Szeg\H{o} projector. In (\ref{uSz}), one can recognize the original {\it cubic Szeg\H{o} equation} as introduced in \cite{GG}. Note that $u(\theta)$ naturally extends to the complex plane by writing $u(z)=\sum_{n=0}^\infty z^n\alpha_n(t)$, in which case $u(z=e^{i\theta})$ recovers the original definition $u(\theta)$. Many statements about the classical system are proved very effectively in this position space picture using properties of Szeg\H{o} projectors, but in preparation for analyzing the quantum system, we will often stress the mode space representation, since, once quantized,
the complex amplitudes $\al_n$ turn into a conventional creation-annihilation operator algebra, while the corresponding field operators $u(\theta)$ acquire rather unusual nonlocal commutation relations.

It was recognized early on \cite{GG} that (\ref{uSz}) is a very special equation and in particular, it admits an infinite hierarchy of dynamically invariant manifolds
formed by meromorphic $u(z)$ with a fixed number of poles in the complex plane. If the initial conditions are of this form, the evolution described by (\ref{uSz}) will
amount to moving the positions of the poles and changing their residues, but the number of singularities will not change. The poles thus act as solitons. In the mode space language, the invariant manifolds are defined by configurations of the form
\beq\label{invM}
\al_n(t)=\sum_{r=1}^R c_r(t)\,[p_r(t)]^n
\eeq
at some prescribed value of $R$.
Further inquiry
into the structure of (\ref{uSz}) has led to the discovery of its Lax pairs that we will briefly summarize immediately below. The original presentation of \cite{GG} revolves around the antilinear operator $H_u h=\Pi(u\bar h)$, resulting in a Lax pair in which the first Lax operator is antilinear. This is used very wisely in the constructions of \cite{GG}, but as it is rather unconventional, we will rely in our presentation on the linear Lax operator obtained by squaring $H_u$.

We proceed with presenting the Lax pair in the mode space language of (\ref{Hcl}). To this end, we introduce auxiliary `test' vectors in the mode space $\vec{h}=(h_0,h_1,\cdots)$, and also, for future use, the special vector
\beq
\vec{1}=(1,0,0,\cdots).
\eeq
The classical Lax pair $(\mathcal{L},\mathcal{M})$ is defined by its action on $\vec{h}$ as
\beq\label{defLMcl}
(\mathcal{L}\vec{h})_n=\sum_{k,l=0}^\infty \al_{n+k}\bar\al_{k+l} h_l ,\qquad (\mathcal{M}\vec{h})_n=\sum_{m+p\ge n}^{m,p\ge 0}\al_p\bar\al_{m+p-n}  h_m.
\eeq
One can show that the relation
\beq\label{LaxLM}
i\dot{\mathcal{L}}=[\mathcal{M},\mathcal{L}]
\eeq
holds whenever the equations of motion (\ref{SzEOM}) are satisfied. (There is an elegant proof of this statement in \cite{GG} using \Sz{} projectors, but for completeness, we provide,  in Appendix~\ref{appLax}, a brute force proof in the mode space language used here.) This has all the usual consequences of the Lax theory, in particular, the conservation of an infinite tower of charges given by $\Tr[\mathcal{L}^p]$ for any $p$, whose time derivatives vanish by (\ref{LaxLM}) and the cyclic property of the trace. But there is much more structure in this particular case than what is dictated by the general Lax theory.

One special property of the Lax operators (\ref{defLMcl}) is 
\beq\label{L1M1}
\mathcal{L}\vec{1}=\mathcal{M}\vec{1}.
\eeq
As a result, one gets an extra Lax pair that provides an extension of the Lax structure (\ref{LaxLM}). Namely, define the projector on $\vec{1}$:
\beq
P_1\vec{h}\equiv (\vec{1},\vec{h})\,\vec{1},
\eeq
where the scalar products are defined by the evident formula
\beq
(\vec{h},\vec{g})=\sum_{n=0}^\infty \bar h_n g_n.
\eeq
As $P_1$ is a time-independent operator, and in view of (\ref{L1M1}), we can write a trivial Lax equation
\beq
i\dot P_1=[\mathcal{M}-\mathcal{L},P_1],
\eeq
where both sides are identically zero. At the same time, we can equivalently rewrite (\ref{LaxLM}) as
\beq
i\dot{\mathcal{L}}=[\mathcal{M}-\mathcal{L},\mathcal{L}]
\eeq
This shows that $\mathcal{L}$ and $P_1$ are compatible Lax operators (with the same Lax partner) and therefore, the trace of any product of powers of $\mathcal{L}$ and $P_1$ will be conserved by the usual Lax construction. In practice, this leads to the conservation of the following quantities:
\beq
I_{n}\equiv\Tr[\mathcal{L}^nP_1]=(\vec{1},\mathcal{L}^n\vec{1})=\hspace{-3mm}\sum_{i_1,\ldots, i_{2n-1}=0}^\infty \hspace{-3mm}\bar\alpha_{i_1} \alpha_{i_1+i_2} \bar\alpha_{i_2+i_3} \alpha_{i_3+i_4}\cdots \bar\alpha_{i_{2n-2}+i_{2n-1}} \alpha_{i_{2n-1}}.\label{In}
\eeq
These coexist with the aforementioned standard Lax conservation laws
\beq
G_n\equiv \Tr[\mathcal{L}^n]=\hspace{-3mm}\sum_{i,i_1,\ldots, i_{2n-1}=0}^\infty \hspace{-3mm}\bar\alpha_{i+i_1} \alpha_{i_1+i_2} \bar\alpha_{i_2+i_3} \alpha_{i_3+i_4}\cdots \bar\alpha_{i_{2n-2}+i_{2n-1}} \alpha_{i_{2n-1}+i},\label{Gn}
\eeq
which however find very little use in the analytic considerations of \cite{GG,GG1,GG2,GG3}. The $I$-tower starts with $I_1=N$, while $I_2$ can be expressed through (\ref{Hcl}) and $N^2$. Similarly, the $G$-tower starts with $G_1=N+M$, while $G_2$ can be related to
$$
\sum_{n+m=k+l}^{n,m,k,l\ge 0} [1+\min(n,m,k,l)]\,\bar\al_n \bar\al_m \al_k \al_l=2H_{cl}+H_{\min,cl} \, ,
$$
with 
\begin{equation}
    H_{\min,cl} \equiv \sum_{n+m=k+l}^{n,m,k,l\ge 0} \min(n,m,k,l)\,\bar\al_n \bar\al_m \al_k \al_l \,,
    \label{eq: Hmin classical}
\end{equation}
the classical analog of the quantum conserved operator $H_{\min}$ defined in (\ref{Hmin}).

There is another Lax-pair construction of a similar sort. Consider $P_\al$ defined by
\beq
P_\al\vec{h}=(\vec{\al},\vec{h})\,\vec{\alpha},
\eeq
where $\vec{\al}=(\al_0,\al_1,\cdots)$. By (\ref{SzEOM}), we have
\beq
i\dot P_\al=[\mathcal{M},P_\al].
\eeq
In other words, $P_\al$ is Lax-compatible with $\mathcal{L}$ (but not with $P_1$). This means we can freely combine $\mathcal{L}$ and $P_\al$, for example, by defining
\beq
\tilde{\mathcal{L}}=\mathcal{L}-P_\al,
\eeq
which would be called $K_u^2$ in the notation of \cite{GG2}. Conservation laws may be built from $\tilde{\mathcal{L}}$, though they are not independent of the conservation laws built from $\mathcal{L}$. One will have the usual traces $\Tr[\tilde{\mathcal{L}}^n]$, as well as
\beq
\tilde I_{n}\equiv\Tr[\tilde{\mathcal{L}}^{n-1}P_\al]=(\vec{\al},\tilde{\mathcal{L}}^{n-1}\vec{\al})=\hspace{-3mm}\sum_{i_1,\ldots, i_{2n-1}=0}^\infty \hspace{-3mm}\bar\alpha_{i_1} \alpha_{i_1+i_2+1} \bar\alpha_{i_2+i_3+1}\cdots \bar\alpha_{i_{2n-2}+i_{2n-1}+1} \alpha_{i_{2n-1}}.\label{I2n}
\eeq

To conclude this brief review of the classical integrability properties of the cubic \Sz{} equation, we mention that a general formula can be written for the evolution of arbitrary prescribed initial configurations. This is much more than what one expects in general from a Lax-integrable system. Translating the original results of \cite{GG2} into the mode space language we are currently using, we introduce the initial data vector $\vec{\al}_{start}\equiv(\al_0(0),\al_1(0),\cdots)$ and the corresponding initial Lax operators $\mathcal{L}_0\equiv \mathcal{L}[\vec{\al}_{start}]$ and $\tilde{\mathcal{L}}_0\equiv \tilde{\mathcal{L}}[\vec{\al}_{start}]$, as well as the shift operator $S$ together with its conjugate $S^\dagger$:
\beq\label{Scl}
S\vec{h}=(0,h_0,h_1,h_2,\cdots),\qquad S^\dagger\vec{h}=(h_1,h_2,h_3,\cdots).
\eeq
Then,
\beq\label{expl}
\al_n(t)=(\vec{1},[e^{-it\mathcal{L}_0}e^{it\tilde{\mathcal{L}}_0}S^\dagger]^ne^{-it\mathcal{L}_0}\vec{\al}_{start}).
\eeq
Further details can be found in \cite{GG2,GG3}.

Lax-integrable deformations of the cubic \Sz{} equation have been constructed \cite{Xu,cascade}, though only part of the analytic structure presented here survives when the equation has been deformed. Finally, we comment on the Lax pair corresponding to the classical evolution defined by \eqref{eq: Hmin classical}, which we found in the course of our investigations. The operator $\mathcal{L}$ of \eqref{defLMcl} remains unchanged, while its Lax partner $\mathcal{M}$ must be replaced with
\begin{equation}
    (\mathcal{M}\vec{h})_n = \sum_{m+p\ge n}^{m,p\ge 0} n \, \al_p\bar\al_{m+p-n}  h_m - \hspace{-0.3cm }\sum^{ \min(n,m)\ge p}_{m,p\ge 0} \hspace{-0.2cm }p \, \al_{n-p}\bar\al_{m-p}  h_m - \hspace{-0.1cm }\sum_{ n+p\ge m}^{m \ge n} (m-n) \, \al_p\bar\al_{n+p-m}  h_m 
\end{equation}
in order to accommodate the evolution generated by the Hamiltonian \eqref{eq: Hmin classical}.


\section{The quantum G\'erard-Grellier Hamiltonian}

We finally return to the quantum Hamiltonian (\ref{HGG}), which will be the main subject of our study for the rest of this paper. As explained at the end of section~\ref{secres}, the space of states of this Hamiltonian is spanned by the Fock vectors
$$
|\eta_0,\eta_1,\cdots\rangle\equiv \left(\prod_{k=0}^\infty \frac{(a^\dagger_k)^{\eta_k}}{\sqrt{\eta_k!}}\right)|0,0,\cdots\rangle,\quad a_k |0,0,\cdots\rangle=0,\quad a_k^\dagger a_k|\eta_0,\eta_1,\cdots\rangle=\eta_k|\eta_0,\eta_1,\cdots\rangle.
$$
Because the operators $N$ and $M$ defined in (\ref{NMq}) commute with the Hamiltonian, the latter can only have nonvanishing matrix elements between the Fock states
with the same values of $N$ (given by $\sum_k\eta_k$) and $M$ (given by $\sum_k k\eta_k$). But since there is only a finite number of such states within each $(N,M)$-block, one is left with independent finite-sized matrices. Diagonalizing these finite-sized matrices reveals integer eigenvalues, providing thereby an entry point \cite{quantres} to the manifold analytic puzzles presented by (\ref{HGG}). What kind of structure can we expect to underlie these patterns, also in view of the classical integrability properties reviewed in section~\ref{seccl}?

The integer eigenvalues suggest that ladder operators must be present that convert eigenvectors to eigenvectors while changing the eigenvalues by integer shifts. We will indeed report some such operators at the end of this section, and more in section~\ref{secladder}.

Quantization of the hierarchies of conservation laws given by (\ref{In}) and (\ref{Gn}) in the classical theory is a very natural outstanding question. Naive quantization
based on replacing $\ab_n$ with $a_n^\dagger$ and $\al_n$ with $a_n$ fails for any obvious ordering prescription. For example, we could try to keep the order as in  (\ref{In}) and (\ref{Gn}), or we could try to impose normal ordering, with all $\ad$'s moved to the left and all $a$'s moved to the right. None of these prescriptions produce valid quantum conservation laws that commute with (\ref{HGG}). In fact, we know that the classical conservation law $G_2$ can be rewritten in terms of a structure identical to the first term of the valid quantum conservation law (\ref{Hmin}). (We will provide a brute force proof that (\ref{Hmin}) commutes with (\ref{HGG}) in section~\ref{secHHmin}.) This makes us expect that other conservation laws will behave similarly: they will consist of the highest-order polynomial term that can be visualized as the corresponding classical conservation law with $(\ab_n,\al_n)$ replaced with $(a_n^\dagger,a_n)$, normal-ordered, plus lower-order polynomial {\it quantum corrections}. This is precisely the structure of (\ref{Hmin}), and it will be true of all other conservation laws we can explicitly construct. What is even more striking is that, for all quantum conservation laws we construct, we empirically find purely {\it integer eigenvalue spectra}. This will be reported in section~\ref{seclax}, together with our knowledge about the rather peculiar quantum Lax pair, also obtained by introducing a quantum correction to the classical Lax pair (\ref{defLMcl}).

Before proceeding with more in-depth studies of the GG Hamiltonian, it will be handy to assemble here a few useful operators and state the algebra they form.
In much of our exposition, we will keep in the back of our mind the picture of diagonalizing simultaneously $H$ given by (\ref{HGG}) and $H_{\min}$ given by (\ref{Hmin}).
First, this will remove some of the possible degeneracies present when $H$ is diagonalized by itself. Second, there are nice algebraic relations that give this simultaneous
diagonalization process a useful purpose.

We first remark that $a_0^\dagger$ commutes with $H_{\min}$, since the latter only involves modes number 1 and higher. For that reason, for any eigenvector $|\Psi\rangle$ of $H_{\min}$ in block $(N,M)$, we can take $a_0^\dagger|\Psi\rangle$ in block $(N+1,M)$ and it will be an eigenvector of $H_{\min}$ as well. Furthermore, we can introduce quantum shift operators $S$ and $S^\dagger$ analogous to (\ref{Scl}):
\beq \label{eq: relations p0 and S}
S|\eta_0,\eta_1,\cdots\rangle=|0,\eta_0,\eta_1,\cdots\rangle,\qquad S^\dagger|\eta_0,\eta_1,\eta_2\cdots\rangle=|\eta_1,\eta_2\cdots\rangle.
\eeq
We define $S^\dagger$ to annihilate $|\eta_0,\eta_1,\cdots\rangle$ unless $\eta_0=0$. With this, we have
\beq
   S^\dagger S = 1,\qquad  SS^\dagger = P_{0}, \qquad P_0S=S,\qquad S^\dagger P_0=S^\dagger,
\eeq
where $P_{0}$ is the projector onto the null space of $a_0$,
as well as the following useful relations:
\begin{align}
    S M S^\dagger &= M-N, \\
    S^\dagger H S &= H, \label{eq:shift1}\\
  H_{\min} S &= S(2H+H_{\min} + 2M+N). \label{eq:shift joint eigenvectors}
\end{align}
The last relation means that, if we have simultaneously diagonalized $H$ and $H_{\min}$ in block $(N,M)$, and $|\Psi\rangle$ is one of their joint eigenvectors, the vector $S|\Psi\rangle$ residing in the block $(N,M+N)$ is an eigenvector of $H_{\min}$ with eigenvalue determined by \eqref{eq:shift joint eigenvectors}.

The above picture suggests a recursive process for diagonalizing $H$ and $H_{\min}$: imagine we have already diagonalized these two operators in the blocks whose values of $N$ or $M$ are below the current block $(N,M)$. 
Then, take the eigenvectors from the block $(k,M-k)$, denoted as $V_{k,M-k}$, and transport them to the block ($k,M$) with $S$. In order to reach the block ($N,M$), one can apply $a_0^\dagger$ acting with it $N-k$ times.
We obtain groups of (mutually) linearly independent vectors, all of which are eigenvectors of $\Hm$. Furthermore, the disjoint union of these vectors
\begin{equation}
    \bigcup_{k=1}^{\min(N,M)} \frac{a_0^{\dagger N-k}}{\sqrt{(N-k)!}} S V_{k,M-k}
    \label{eq: construction Hmin eigenspace}
\end{equation}
forms a complete basis in the block $(N,M)$. This follows directly from the linear independence of these states, combined with a counting argument relating the dimensions of the spaces $V_{k,M-k}$ and the size of the block $(N,M)$. The number of vectors in an ($N,M$) block is the number of integer partitions of $M$ into at most $N$ parts \cite{quantres} denoted $p_N(M)$. This number-theoretic function satisfies the recursion relation
\beq
p_N(M)=p_{N-1}(M)+p_N(M-N).
\label{eq: pnm relation}
\eeq
(A partition of $M$ into at most $N$ parts is either a partition into exactly $N$ parts, so that no parts are zero and we can subtract 1 from each part, obtaining a partition of $M-N$ into at most $N$ parts, or it is a partition of $M$ into at most $N-1$ part.) Applying this relation recursively to the first term on the right-hand side of \eqref{eq: pnm relation} yields
\beq
p_N(M)= \sum_{k=1}^{\min(N,M)}p_{k}(M-k) \, .
\eeq
This shows that \eqref{eq: construction Hmin eigenspace} contains precisely enough linearly independent vectors to be a basis for the block $(N,M)$.
We have thus obtained, by this very simple process, an eigenbasis of $H_{\min}$ in the current block. The only issue is that, to restart the iterative procedure we need a joint eigenbasis of $H$ and $H_{\min}$, which may require re-diagonalization within the degenerate subspaces of $H_{\min}$. So we must inspect the spectrum and rediagonalize $H$ within any such degenerate subspaces.

We will see in sections~\ref{secHmintop} and \ref{secHmintop2} that, in application to the two highest subspaces of $H_{\min}$, this procedure gives an explicit construction of families of eigenvectors. The entry point of this construction is the joint top eigenvectors of $H$ and $H_{\min}$ that we will build in section~\ref{sechighest}. There is a relation between these top eigenspaces of $H_{\min}$ and classical invariant manifolds (\ref{invM}), as we shall explain in section~\ref{secinvman}. Before proceeding with that, we will construct, in section~\ref{secbound}, bounds on $H$ and $H_{\min}$ that will help us identify these top subspaces. For future use we define, in addition to the operators introduced above, the following family of operators
\beq
J_1\equiv a_0,\qquad J_{2n+1}=[J_{2n-1},H].
\label{oddJs}
\eeq
These are quantum analogs of the classical quantities with the same name in \cite{GG}.


\section{Energy bounds}\label{secbound}

The bounds on $H$ and $H_{\min}$ will play an essential role in our constructions of explicit families of eigenvectors below. They are also important more broadly in
controlling the spectral properties of $H$ and $H_{\min}$, and will reveal useful algebraic structures.
Consider first the following decomposition:
\begin{align}\label{boundH}
(N-1)&(N+2M)-2H=\sum_{kl}(1+k+l)\ad_k\ad_l a_ka_l-2H \nonumber \\
&=\sum_{j=0}^\infty\left[(1+j)\sum_{k=0}^j \ad_k\ad_{j-k} a_ka_{j-k}-\sum_{k,l=0}^j \ad_k\ad_{j-k} a_la_{j-l} \right]
=\sum_{j=0}^\infty\sum_{\alpha=1}^{\lfloor{j/2}\rfloor} B_{j\alpha}^\dagger B_{j\alpha},
\end{align}
where
\beq
B_{j\alpha}\equiv \sum_{k=0}^j e^{(j,\alpha)}_k a_k a_{j-k},
\eeq
with $e^{(j,\alpha)}_k$ being any set of vectors satisfying
\beq \label{eq:conditions ek}
e^{(j,\alpha)}_k=e^{(j,\alpha)}_{j-k},\qquad \sum_k e^{(j,\alpha)}_k=0,\qquad \sum_{\alpha=1}^{\lfloor{j/2}\rfloor} e^{(j,\alpha)}_ke^{(j,\alpha)}_l=(1+j)\frac{(\delta_{kl}+\delta_{k,j-l})}{2}-1.
\eeq
(One can check that the bisymmetric matrix appearing on the right-hand side is semi-positive-definite and admits the stated decomposition.)
Since the right-hand side of (\ref{boundH}) is evidently positive-definite, we get the bound
\beq
2H\le (N-1)(N+2M).
\eeq
This bound is saturated on a vector $|\Psi\rangle$ if and only if
\beq
\forall j,\alpha:\quad B_{j\alpha}|\Psi\rangle=0.
\eeq
We will use this and other similar formulas below to construct specific families of eigenvectors. Note that $(N-1)(N+2M)$ is nothing but the normal-ordered product of $N$ and $N+2M$.

The Hamiltonian $H_{\min}$ may be visualized in terms of a `layered cake' construction: we first take $H$ and shift all the mode indices of the $a$'s by +1, then by +2, +3 and so on, and add up the results, obtaining
\beq\label{layer}
H_{\min} = \sum_{n+m=k+l}^{n,m,k,l\ge 0} \sum_{s=1}^\infty a^\dagger_{n+s} a^\dagger_{m+s} a_{k+s} a_{l+s} +  \sum_{k=1}^\infty k^2 a^\dagger_k a_k.
\eeq
Indeed, a given term $a^\dagger_{\bar n} a^\dagger_{\bar m} a_{\bar k} a_{\bar l}$ will be present in the above sum for $s=1..\min(\bar n,\bar m,\bar k, \bar l)$, yielding the coefficient in (\ref{Hmin}) upon summation over $s$.
We can then apply the decomposition (\ref{boundH}) within this `layered cake' construction:
\beq
\sum_{n+m=k+l}^{n,m,k,l\ge 0} a^\dagger_{n+s} a^\dagger_{m+s} a_{k+s} a_{l+s} =\sum_{j=0}^\infty(1+j)\sum_{k=0}^j \ad_{k+s}\ad_{j-k+s} a_{k+s}a_{j-k+s}-\sum_{j=0}^\infty\sum_{\alpha=1}^{\lfloor{j/2}\rfloor} B_{js\alpha}^\dagger B_{js\alpha},
\eeq
where
\beq
B_{js\alpha}\equiv \sum_{k=0}^{j} e^{(j,\alpha)}_k a_{k+s} a_{j-k+s}.
\eeq
Now,
\begin{align}
 &\sum_{s=1}^\infty\sum_{j=0}^\infty(1+j)\sum_{k=0}^j \ad_{k+s}\ad_{j-k+s} a_{k+s}a_{j-k+s}= \sum_{s=1}^\infty\sum_{k,l=0}^\infty(1+k+l) \ad_{k+s}\ad_{l+s} a_{k+s}a_{l+s}\\
&=  \sum_{s=1}^\infty\sum_{k,l=s}^\infty(1+k+l-2s) \ad_{k}\ad_{l} a_{k}a_{l}=\sum_{k,l=1}^\infty \ad_{k}\ad_{l} a_{k}a_{l}\sum_{s=1}^{\min(k,l)}(1+k+l-2s).
\end{align}
For the last sum, we get
\begin{align}
\sum_{s=1}^{\min(k,l)}(1+k+l-2s)&=(1+k+l)\min(k,l)-(\min(k,l)+1)\min(k,l)\\
&=[k+l-\min(k,l)]\min(k,l)=\max(k,l)\min(k,l)=kl.\nonumber
\end{align}
 Furthermore,
\beq
\sum_{k,l=1}^\infty kl\,\ad_{k}\ad_{l} a_{k}a_{l}+\sum_{k=1}^\infty k^2\ad_{k}a_{k}=M^2
\eeq
Putting everything together,
\beq\label{minbound}
H_{\min}=M^2-\sum_{s=1}^\infty\sum_{j=0}^\infty\sum_{\alpha=1}^{\lfloor{j/2}\rfloor} B_{js\alpha}^\dagger B_{js\alpha}.
\eeq

Not all of the conditions $B_{js\alpha}|\Psi\rangle=0$ are independent. For example, enforcing this condition for $s=1$ and all $j$ and $\alpha$ automatically enforces it for all higher values of $s$. For such vectors $H_{\min}=M^2$. If, on the other hand, the conditions hold for $s=2$ but not for $s=1$, they hold automatically for all $s\ge 2$, and only terms with $s=1$ are left in (\ref{minbound}). These terms can furthermore be simplified using $B_{js\alpha}|\Psi\rangle=0$ with $s\ge 2$ since that should eliminate subsets of terms in $B_{j1\alpha}$. We will use such subspaces defined by $B_{js\alpha}|\Psi\rangle=0$ for diagonalizing $H$ and $H_{\min}$.

In (\ref{boundH}), we used a complete-basis decompositions of quadratic forms, ending up with somewhat awkward expressions in terms of the eigenvectors $e_k$. In many ways, it is more convenient to take advantage of the formula
\beq
(1+j)\sum_{k=0}^j\bar x_k x_k-\sum_{k,l=0}^j \bar x_k x_l= \frac12 \sum_{k,l=0}^j (\bar x_k x_k+\bar x_l x_l-\bar x_k x_l-\bar x_l x_k)=
\frac12 \sum_{k,l=0}^j |x_k-x_l|^2,
\eeq
and write instead of (\ref{boundH})
\beq\label{boundHD}
2H=(N-1)(N+2M)-\frac12\sum_{j=0}^\infty\sum_{k,l=0}^j(a_k a_{j-k}-a_l a_{j-l})^\dagger (a_k a_{j-k}-a_l a_{j-l}).
\eeq
An analogous representation of the bound (\ref{minbound}) is
\beq \label{eq:HminForConstraints}
H_{\min} =M^2-\frac12\sum_{s=1}^\infty\sum_{j=0}^\infty\sum_{k,l=0}^j(a_{k+s} a_{j-k+s}-a_{l+s} a_{j-l+s})^\dagger (a_{k+s} a_{j-k+s}-a_{l+s} a_{j-l+s}).
\eeq
Introducing $\tilde k=k+s$, $\tilde l =l+s$, $\tilde j=j+2s$ and dropping tildes, we get
\beq\label{boundHminD}
H_{\min} =M^2-\frac12\sum_{s=1}^\infty\sum_{j=2s}^\infty\sum_{k,l=s}^{j-s}(a_{k} a_{j-k}-a_{l} a_{j-l})^\dagger (a_{k} a_{j-k}-a_{l} a_{j-l}).
\eeq
The summand is now $s$-independent, so the sums can be simplified. The index range conditions $s\le k\le j-s$ and $s\le l\le j-s$ can be recast as $s\le k$, $s\le j-k$, $s\le l$, $s\le j-l$ or, equivalently,
$s\le \min(k,j-k,l,j-l)$, and then
\beq\label{boundHmD}
H_{\min} =M^2-\frac12\sum_{j=0}^\infty\sum_{k,l=0}^{j}\min(k,j-k,l,j-l)(a_{k} a_{j-k}-a_{l} a_{j-l})^\dagger (a_{k} a_{j-k}-a_{l} a_{j-l}).
\eeq

It will be economical for some purposes to switch to thinking in terms of the following two Hamiltonians:
\beq\label{H0}
H_0=\frac12\sum_{j=0}^\infty\sum_{k,l=0}^j D^\dagger_{jkl}D_{jkl},
\eeq
and
\beq\label{H1}
H_1=\frac12\sum_{j=0}^\infty\sum_{k,l=0}^j \min(k,j-k,l,j-l)D^\dagger_{jkl}D_{jkl},
\eeq
with
\beq
D_{jkl}\equiv a_{k} a_{j-k}-a_{l} a_{j-l}.
\eeq
Evidently, 
\beq\label{HHmH0H1}
2H=(N-1)(N+2M)-H_0,\qquad H_{\min}=M^2-H_1.
\eeq 
An advantage is that $H_1$ is purely quartic, in contrast with (\ref{Hmin}) that has a quadratic `quantum' piece. Viewed from the vantage point of the above formulas, this quadratic piece has a trivial origin: it comes from normal-ordering $M^2$. We will rely on $H_0$ and $H_1$ in the next section, and prove that they commute, which will automatically imply that $[H,H_{\min}]=0$.

We remark that sum-of-squares decompositions like (\ref{H0}) and (\ref{H1}) are very typical of the `factorization method' and supersymmetric quantum mechanics \cite{factorization}. However, since $D$ and $D^\dagger$ do not commute for different index values, it is not immediately obvious how to use them for constructing a full solution. 


\section{A brute force proof that $[H,H_{\min}]=0$}\label{secHHmin}

Since we use simultaneous diagonalization of $H$ and $\Hm$ in our constructions of the eigenvectors below, the property $[H,\Hm]=0$ is essential for our considerations.
The Hamiltonian $\Hm$ was originally guessed in \cite{bound} as an outcome of a numerical procedure that algorithmically constructs conservation laws polynomial in the creation-annihilation operators as explicit matrices within the individual $(N,M)$-blocks of the Hilbert space. It is easy to verify within any given $(N,M)$-block that the two operators indeed commute, but what we need is a general analytic proof.

While we expect that, eventually, a systematic understanding of the conservation laws of $H$ will emerge in the spirit of Lax theory, we are far from that point, and a practical solution is to provide a brute force proof that  $[H,\Hm]=0$. It is more convenient to prove instead that
$[H_0,H_1]=0$, with the definitions \eqref{H0} and \eqref{H1}, which would automatically imply $[H,\Hm]=0$ in view of (\ref{HHmH0H1}). We provide a proof below. It is
essential for our subsequent considerations, but rather bulky and tedious. The rest of our treatment will only use the final result, but not the details of this proof, so the details can be freely skipped.

We remark that we could in principle throw away the top polynomial piece of $[H_0,H_1]$ after normal ordering, because that piece must agree with the classical computation, and in the classical theory, $H$ and $H_{\min}$ Poisson-commute from the general Lax theory described in section~\ref{seccl}. We opt, however, for a completely mechanical proof that follows through detailed bookkeeping of all the terms emerging from the commutation and proceeds to demonstrate that the result is zero.

We introduce the shorthand $\sum_{\{jkl\}}\equiv\sum_{j=0}^\infty\sum_{k,l=0}^j$. Then,
\beq
[H_0,H_1]=\frac14\sum_{\{jkl\}}\sum_{\{j'k'l'\}} \mathrm{min}_{j'k'l'}[D^\dagger_{jkl}D_{jkl},D^\dagger_{j'k'l'}D_{j'k'l'}],
\eeq
with the definition
\beq
\mathrm{min}_{jkl}\equiv\min(k,j-k,l,j-l).
\eeq
We have
\begin{align*}
[D^\dagger_{jkl}D_{jkl},D^\dagger_{j'k'l'}D_{j'k'l'}]&=D^\dagger_{jkl}D_{jkl}D^\dagger_{j'k'l'}D_{j'k'l'}-D^\dagger_{j'k'l'}D_{j'k'l'}D^\dagger_{jkl}D_{jkl}\\
&=D^\dagger_{jkl}[D_{jkl},D^\dagger_{j'k'l'}]D_{j'k'l'}-D^\dagger_{j'k'l'}[D_{j'k'l'},D^\dagger_{jkl}]D_{jkl}.
\end{align*}
So,
$$
[H_0,H_1]=\sum_{\{jkl\}}\sum_{\{j'k'l'\}}\mathrm{min}_{j'k'l'}
\left(D^\dagger_{jkl}[a_ka_{j-k},\ad_{k'}\ad_{j'-k'}]D_{j'k'l'}-D^\dagger_{j'k'l'}[a_{k'}a_{j'-k'},\ad_{k}\ad_{j-k}]D_{jkl}\right),
$$
where we have used the antisymmetry of $D_{jkl}$ with respect to permuting $k$ and $l$. Now,
\begin{align*}
[a_ka_{j-k},\ad_{k'}\ad_{j'-k'}]&=\de_{kk'}a_{j-k}\ad_{j'-k'}+\de_{k,j'-k'}a_{j-k}\ad_{k'}+\de_{j-k,k'}a_{k}\ad_{j'-k'}+\de_{j-k,j'-k'}a_k\ad_{k'}\\
&=\de_{kk'}\ad_{j'-k'}a_{j-k}+\de_{k,j'-k'}\ad_{k'}a_{j-k}+\de_{j-k,k'}\ad_{j'-k'}a_{k}+\de_{j-k,j'-k'}\ad_{k'}a_k\\
&+2\de_{kk'}\de_{jj'}+2\de_{k,j-k'}\de_{jj'}.
\end{align*}
Plugging this back in and keeping in mind that $j-k$ can be interchanged with $k$ using the symmetry of $D_{jkl}$, and similarly for the primed indices, we get
\beq\label{H0H1}
\begin{split}
\frac{[H_0,H_1]}4&=\sum_{\{jkl\}}\sum_{\{j'k'l'\}} \mathrm{min}_{j'k'l'}
\big[D^\dagger_{jkl}(\de_{kk'}\ad_{j'-k'}a_{j-k}+\de_{kk'}\de_{jj'})D_{j'k'l'}\\
&\hspace{3cm}-D^\dagger_{j'k'l'}(\de_{kk'}\ad_{j-k}a_{j'-k'}+\de_{kk'}\de_{jj'})D_{jkl}\big]\\
&=\sum_{j,j'=0}^\infty\sum_{l=0}^j\sum_{l'=0}^{j'}\sum_{k=0}^{\min(j,j')} \mathrm{min}_{j'kl'}\big[D^\dagger_{jkl}\ad_{j'-k}a_{j-k}D_{j'kl'}-(jl\leftrightarrow j'l')\big]\\
&\hspace{2cm}+\sum_{j=0}^\infty\sum_{k,l,l'=0}^j \mathrm{min}_{jkl'}\big[D^\dagger_{jkl}D_{jkl'}-D^\dagger_{jkl'}D_{jkl}\big].
\end{split}
\eeq
Take the last line:
\begin{align*}
&\sum_{j=0}^\infty\sum_{k,l,l'=0}^j \mathrm{min}_{jkl'}\big[D^\dagger_{jkl}D_{jkl'}-D^\dagger_{jkl'}D_{jkl}\big]=\sum_{j=0}^\infty\sum_{k,l,l'=0}^j\big[ \mathrm{min}_{jkl'}- \mathrm{min}_{jkl}\big]D^\dagger_{jkl}D_{jkl'}\\
&=\sum_{j=0}^\infty\sum_{k,l,l'=0}^j\big[ \mathrm{min}_{jkl'}- \mathrm{min}_{jkl}\big]\big(\ad_k\ad_{j-k}a_ka_{j-k}-\ad_l\ad_{j-l}a_ka_{j-k}-\ad_k\ad_{j-k}a_{l'}a_{j-l'}+\ad_l\ad_{j-l}a_{l'}a_{j-l'}\big).
\end{align*}
The first term in the round brackets does not depend on $l$ and $l'$, while $\sum_{ll'}(\mathrm{min}_{jkl'}- \mathrm{min}_{jkl})$ is evidently 0, thereby eliminating the first term in the round brackets. For the second term in the round brackets, we do the substitution $k\to l'\to l\to k$, which gives
$$
\sum_{j=0}^\infty\sum_{k,l,l'=0}^j\big[ \mathrm{min}_{jll'}- \mathrm{min}_{jkl'}\big]\ad_k\ad_{j-k}a_{l'}a_{j-l'}.
$$
This combines nicely with the third term in the round brackets, leaving 
\begin{align*}
&\sum_{j=0}^\infty\sum_{k,l,l'=0}^j \mathrm{min}_{jkl'}\big[D^\dagger_{jkl}D_{jkl'}-D^\dagger_{jkl'}D_{jkl}\big]\\
&=-\sum_{j=0}^\infty\sum_{k,l,l'=0}^j\big[ \mathrm{min}_{jll'}- \mathrm{min}_{jkl}\big]\ad_k\ad_{j-k}a_{l'}a_{j-l'}
+\sum_{j=0}^\infty\sum_{k,l,l'=0}^j\big[ \mathrm{min}_{jkl'}- \mathrm{min}_{jkl}]\ad_l\ad_{j-l}a_{l'}a_{j-l'}.
\end{align*}
Finally, we interchange $k$ and $l$ in the first sum, making the two sums cancel each other. We have thus proved that the last line in (\ref{H0H1}) vanishes. Note that, up to this point, we have not used the explicit form of $\mathrm{min}_{jkl}$ but only its index permutation symmetries.

We now turn to the first line of the last representation in (\ref{H0H1}):
\begin{align}
&\sum_{j,j'=0}^\infty\sum_{l=0}^j\sum_{l'=0}^{j'}\sum_{k=0}^{\min(j,j')} (\mathrm{min}_{j'kl'}-\mathrm{min}_{jkl})D^\dagger_{jkl}\ad_{j'-k}a_{j-k}D_{j'kl'}\nn\\
&=\sum_{j,j'=0}^\infty\sum_{l=0}^j\sum_{l'=0}^{j'}\sum_{k=0}^{\min(j,j')}  (\mathrm{min}_{j'kl'}-\mathrm{min}_{jkl})
(\ad_{k}\ad_{j-k}\ad_{j'-k}a_{j-k}a_{k}a_{j'-k}
-\ad_{k}\ad_{j-k}\ad_{j'-k}a_{j-k}a_{l'}a_{j'-l'}\nn\\
&\hspace{5cm}
-\ad_{l}\ad_{j-l}\ad_{j'-k}a_{j-k}a_{k}a_{j'-k}
+\ad_{l}\ad_{j-l}\ad_{j'-k}a_{j-k}a_{l'}a_{j'-l'})\nn\\
&=\sum_{k=0}^{\infty} \sum_{j,j'=k}^\infty\sum_{l=0}^j\sum_{l'=0}^{j'} (\mathrm{min}_{j'kl'}-\mathrm{min}_{jkl})
(\ad_{k}\ad_{j-k}\ad_{j'-k}a_{j-k}a_{k}a_{j'-k}
-\ad_{k}\ad_{j-k}\ad_{j'-k}a_{j-k}a_{l'}a_{j'-l'}\nn\\
&\hspace{5cm}
-\ad_{l}\ad_{j-l}\ad_{j'-k}a_{j-k}a_{k}a_{j'-k}
+\ad_{l}\ad_{j-l}\ad_{j'-k}a_{j-k}a_{l'}a_{j'-l'})\nn\\
&=\sum_{k,j,j'=0}^{\infty} \sum_{l=0}^{j+k}\sum_{l'=0}^{j'+k} (\mathrm{min}_{j'+k,kl'}-\mathrm{min}_{j+k,kl})
[\ad_{k}\ad_{j}\ad_{j'}a_{k}a_{j}a_{j'}
-\ad_{k}\ad_{j}\ad_{j'}a_{l'}a_{j}a_{j'+k-l'}\nn\\
&\hspace{5cm}
-\ad_{l}\ad_{j'}\ad_{j+k-l}a_{k}a_{j}a_{j'}
+\ad_{l}\ad_{j'}\ad_{j+k-l}a_{l'}a_{j}a_{j'+k-l'}].\label{H0H1quart}
\end{align}
The first term in the square brackets vanishes using the permutation $jl\leftrightarrow j'l'$. Then, take the third term:
\begin{align*}
&\sum_{k,j,j'=0}^{\infty} \sum_{l=0}^{j+k}\sum_{l'=0}^{j'+k} (\mathrm{min}_{j'+k,kl'}-\mathrm{min}_{j+k,kl})
\ad_{l}\ad_{j'}\ad_{j+k-l}a_{k}a_{j}a_{j'}\\
&=\sum_{j,j',l=0}^{\infty} \sum_{k=\max(0,l-j)}^\infty\sum_{l'=0}^{j'+k} (\mathrm{min}_{j'+k,kl'}-\mathrm{min}_{j+k,kl})
\ad_{l}\ad_{j'}\ad_{j+k-l}a_{k}a_{j}a_{j'}\\
&=\sum_{j,j',l=0}^{\infty} \sum_{k=\max(j-l,0)}^\infty\sum_{l'=0}^{j'+k+l-j} (\mathrm{min}_{j'+k+l-j,k+l-j,l'}-\mathrm{min}_{l+k,k+l-j,l})
\ad_{l}\ad_{j'}\ad_{k}a_{k+l-j}a_{j}a_{j'}\\
&=\sum_{k,l,j'=0}^{\infty} \sum_{j=0}^{k+l}\sum_{l'=0}^{j'+k+l-j} (\mathrm{min}_{j'+k+l-j,k+l-j,l'}-\mathrm{min}_{l+k,k+l-j,l})
\ad_{l}\ad_{j'}\ad_{k}a_{k+l-j}a_{j}a_{j'}\\
&=\sum_{k,j,j'=0}^{\infty} \sum_{l=0}^{j+k}\sum_{l'=0}^{j+j'+k-l} (\mathrm{min}_{j+j'+k-l,j+k-l,l'}-\mathrm{min}_{j+k,j+k-l,j})
\ad_{k}\ad_{j}\ad_{j'}a_{j+k-l}a_{l}a_{j'}\\
&=\sum_{k,j,j'=0}^{\infty} \sum_{l=0}^{j+k}\sum_{l'=0}^{j+j'+k-l} (\mathrm{min}_{j+j'+k-l,l'j'}-\mathrm{min}_{j+k,kl})
\ad_{k}\ad_{j}\ad_{j'}a_{j+k-l}a_{l}a_{j'}\\
&=\sum_{k,j,j'=0}^{\infty} \sum_{l=0}^{j+k}\sum_{l'=0}^{j'+l} (\mathrm{min}_{j'+l,ll'}-\mathrm{min}_{j+k,kl})
\ad_{k}\ad_{j}\ad_{j'}a_{j+k-l}a_{l}a_{j'}\\
&=\sum_{k,j,j'=0}^{\infty} \sum_{l'=0}^{j'+k}\sum_{l=0}^{j+l'} (\mathrm{min}_{j+l',ll'}-\mathrm{min}_{j'+k,kl'})
\ad_{k}\ad_{j}\ad_{j'}a_{l'}a_{j}a_{j'+k-l'}\\
&=\sum_{k,j,j'=0}^{\infty} \sum_{l'=0}^{j'+k}(jl'-(j+l'+1)\mathrm{min}_{j'+k,kl'})
\ad_{k}\ad_{j}\ad_{j'}a_{l'}a_{j}a_{j'+k-l'}\\
&=\frac12\sum_{k,j,j'=0}^{\infty} \sum_{l'=0}^{j'+k}[j(j'+k)-(2j+j'+k+2)\mathrm{min}_{j'+k,kl'}]
\ad_{k}\ad_{j}\ad_{j'}a_{l'}a_{j}a_{j'+k-l'},
\end{align*}
where we have used
\beq
\sum_{k=0}^j\mathrm{min}_{jkl}=\sum_{k=0}^j\min(k,j-k,l,j-l)=l(j-l),
\eeq
and then symmetrized with respect to $l'\leftrightarrow j'+k-l'$.
The second term in (\ref{H0H1quart}) gives by direct evaluation
\begin{align*}
&\sum_{k,j,j'=0}^{\infty} \sum_{l=0}^{j+k}\sum_{l'=0}^{j'+k} (\mathrm{min}_{j'+k,kl'}-\mathrm{min}_{j+k,kl})
\ad_{k}\ad_{j}\ad_{j'}a_{l'}a_{j}a_{j'+k-l'}\\
&=\sum_{k,j,j'=0}^{\infty} \sum_{l'=0}^{j'+k} [(j+k+1)\mathrm{min}_{j'+k,kl'}-jk]
\ad_{k}\ad_{j}\ad_{j'}a_{l'}a_{j}a_{j'+k-l'}.
\end{align*}
Adding this up with the third term we get,
\beq
\frac12\sum_{k,j,j'=0}^{\infty} \sum_{l'=0}^{j'+k}[j(j'-k)-(j'-k)\mathrm{min}_{j'+k,kl'}]
\ad_{k}\ad_{j}\ad_{j'}a_{l'}a_{j}a_{j'+k-l'},
\eeq
but this expression equal minus itself under the interchange of $k$ and $j'$, and hence the sum of the second and third terms in (\ref{H0H1quart}) vanishes. 

Finally, take the last term in (\ref{H0H1quart}) and process it in a manner similar to the third term, so that the indices of the three $\ad$'s are $k$, $j$ and $j'$: 
\begin{align*}
&\sum_{k,j,j'=0}^{\infty} \sum_{l=0}^{j+k}\sum_{l'=0}^{j'+k} (\mathrm{min}_{j'+k,kl'}-\mathrm{min}_{j+k,kl})\ad_{l}\ad_{j'}\ad_{j+k-l}a_{l'}a_{j}a_{j'+k-l'}\\
&=\sum_{j,j',l=0}^{\infty} \sum_{k=\max(0,l-j)}^\infty\sum_{l'=0}^{j'+k} (\mathrm{min}_{j'+k,kl'}-\mathrm{min}_{j+k,kl})\ad_{l}\ad_{j'}\ad_{j+k-l}a_{l'}a_{j}a_{j'+k-l'}\\
&=\sum_{j,j',l=0}^{\infty} \sum_{k=\max(j-l,0)}^\infty\sum_{l'=0}^{j'+k+l-j} (\mathrm{min}_{j'+k+l-j,k+l-j,l'}-\mathrm{min}_{k+l,k+l-j,l})\ad_{l}\ad_{j'}\ad_{k}a_{l'}a_{j}a_{j'+l+k-j-l'}\\
&=\sum_{k,l,j'=0}^{\infty} \sum_{j=0}^{l+k}\sum_{l'=0}^{j'+k+l-j} (\mathrm{min}_{j'+k+l-j,k+l-j,l'}-\mathrm{min}_{k+l,k+l-j,l})\ad_{l}\ad_{j'}\ad_{k}a_{l'}a_{j}a_{j'+l+k-j-l'}\\
&=\sum_{k,j,j'=0}^{\infty} \ad_{j}\ad_{j'}\ad_{k}\sum_{l=0}^{j+k}\sum_{l'=0}^{j+j'+k-l} (\mathrm{min}_{j+j'+k-l,j+k-l,l'}-\mathrm{min}_{j+k,j+k-l,j})a_{l}a_{l'}a_{j+j'+k-l-l'}\\
&=\sum_{k,j,j'=0}^{\infty} \ad_{j}\ad_{j'}\ad_{k}\sum_{l=0}^{j+k}\sum_{l'=0}^{j+j'+k-l} (\mathrm{min}_{j+j'+k-l,j'l'}-\mathrm{min}_{j+k,lk})a_{l}a_{l'}a_{j+j'+k-l-l'}\\
&\equiv\sum_{k,j,j'=0}^{\infty} \ad_{j}\ad_{j'}\ad_{k}A_{kjj'}=\frac13\sum_{k,j,j'=0}^{\infty} \ad_{j}\ad_{j'}\ad_{k}(A_{kjj'}+A_{kj'j}+A_{j'jk}),
\end{align*}
with
\beq\label{Adef}
A_{kjj'}\equiv\sum_{l=0}^{j+k}\sum_{l'=0}^{j'+l} (\mathrm{min}_{j'+l,ll'}-\mathrm{min}_{j+k,lk})a_{j+k-l}a_{l'}a_{j'+l-l'}.
\eeq
One can easily verify using computer algebra for any specific $k$, $j$ and $j'$ that $A_{kjj'}+A_{kj'j}+A_{j'jk}=0$, but proving this directly appears challenging because of the complexity of the summation regions. This difficulty can be bypassed by converting sums into integrals using generating functions, and we present a complete proof in Appendix~\ref{apphhmin}. With this, we have established that the last term in the square brackets of (\ref{H0H1quart}) gives a vanishing contribution, while all the other terms have been proved earlier to give vanishing contributions as well, hence $[H_0,H_1]=0$, and therefore $[H,H_{\min}]=0$.


\section{The highest energy states}\label{sechighest}

Each $(N,M)$-block contains one eigenvector with the highest possible eigenvalues of $H$ and $\Hm$ defined by the bounds (\ref{boundHD}) and (\ref{boundHmD}). These eigenvalues are $(N-1)(N+2M)/2$ and $M^2$, respectively. The corresponding vector $|\psi^{\max}_{N,M}\rangle$ must satisfy, by (\ref{boundHD}) and (\ref{boundHmD}),
\beq\label{eq: constraints highest energy state} 
 a_{k} a_{j-k}|\psi^{\max}_{N,M}\rangle= a_{l} a_{j-l}|\psi^{\max}_{N,M}\rangle
\eeq
for all $j$, $k$ and $l$. We note that the constraints \eqref{eq: constraints highest energy state} are preserved by the action of the following operator
\begin{equation}
    Z = \sum_{n=0}^\infty (n+1) a^{\dagger}_{n+1}a_n \, ,
    \label{eq:Zoperator}
\end{equation}
which moves states from the block $(N,M)$ to $(N,M+1)$. If we assume that $\ket{\psi}$ satisfies \eqref{eq: constraints highest energy state}, then $Z\ket{\psi}$ also satisfies \eqref{eq: constraints highest energy state}: 
\begin{align}
    a_{j-k} a_k Z |\psi \rangle &= Z a_{j-k} a_k |\psi \rangle + \left[(1-\delta_{k0})k a_{j-k} a_{k-1}   + (1-\delta_{j-k,0})(j-k) a_{j-k-1} a_k\right]|\psi \rangle \nonumber \\
				&= Z a_{j-l} a_l |\psi \rangle + j a_{j-1} a_0 |\psi \rangle 
				= a_{j-l} a_l Z |\psi \rangle + (j a_{j-1} a_0 - j a_{j-1} a_0 ) |\psi \rangle = a_{j-l} a_l Z |\psi \rangle.\nn
\end{align}

A possible strategy is therefore to solve the constraints \eqref{eq: constraints highest energy state} in low $M$ blocks and use the $Z$-operator to transport the vectors to higher $M$ blocks. For this purpose, we consider the blocks with $M=0$, which contain a single state $| N,0,...\rangle$. Such a state evidently satisfies \eqref{eq: constraints highest energy state}. Hence, at least one state that saturates the bounds  (\ref{boundHD}) and (\ref{boundHmD}) can be found in each $(N,M)$ block by transporting the single state in block ($N,0$) to the block $(N,M)$ by the repeated action of the $Z$-operator:
\begin{equation}
    \ket{\psi^{\max}_{N,M}} = \frac{Z^{  M}| N,0,\cdots\rangle}{\sqrt{ \langle N,0,\cdots | Z^{\dagger  M}Z^{  M}| N,0,...\rangle}}
    \label{eq: highest energy vector}
\end{equation}
Note that the kernel of $Z$ is trivial.\footnote{The $Z$-operator has appeared in our previous resonant models studies. Specifically, it corresponds to the raising operator in \cite{qperiod2} with $\delta = 1$. In that work, it was shown that the $Z$-operator does not annihilate states.} Therefore, the states \eqref{eq: highest energy vector} define genuine eigenstates of $H$ and $\Hm$. The corresponding eigenvalues are $(N-1)(N+2M)/2$ and $M^2$, since they saturate \eqref{boundHD} and \eqref{boundHmD}.
The denominator in \eqref{eq: highest energy vector} is relatively straightforward to compute using the identity \cite{qperiod2}
\begin{equation}
Z^{\dagger m} Z^{m} = \sum_{k=0}^{m} \frac{m!^2}{k! (m-k)!^2} \frac{(k-1+ 2\hat{M} +\hat{N})!}{(-1+ 2\hat{M} +\hat{N})!} Z^{m-k}Z^{\dagger m-k}.
\label{eq: commutators of Zs}
\end{equation}
Acting with this on the state $| N,0,...\rangle$, the only term that contributes is $k=m$, such that 
\begin{equation}
    \ket{{\psi}^{\max}_{N,M}} = \sqrt{\frac{(N-1)!}{M!(N+M-1)!}}Z^{  M}| N,0,...\rangle .
     \label{eq: highest energy state final}
\end{equation}

It turns out that the bounds \eqref{boundHD} and \eqref{boundHmD} that lead to \eqref{eq: constraints highest energy state} are saturated by a single state in each $(N,M)$ block, given by \eqref{eq: highest energy state final}. The proof for uniqueness relies on showing that all the Fock basis coefficients of a state satisfying \eqref{eq: constraints highest energy state} are related, so that $\ket{{\psi}^{\max}_{N,M}}$ is unique up to normalization. Take any two Fock basis vectors $\ket{e_1}$ and $\ket{e_2}$. There exist two products of annihilation operators $L_1$ and $L_2$ that turn them into the same vector (up to normalization) $L_1\ket{e_1} \sim L_2\ket{e_2}$ in some lower $(N',M')$-block. Since both vectors started in the block $(N,M)$ and have been mapped to the block $(N',M')$, the number of annihilation operators and the sum of their indices must be equal in both $L_1$ and $L_2$. The claim is that for a vector $\ket{\psi}$ satisfying \eqref{eq: constraints highest energy state}, one has $L_1\ket{\psi} = L_2 \ket{\psi}$. A constructive approach to verify this claim, is to consider extreme values $a_{j-k}a_k$ (with $j-k>k$) that are present in one of the $L_i$ but not in both, and exchange them for $a_{j-k-1}a_{k+1}$ using \eqref{eq: constraints highest energy state}. One can repeat this on either side, depending on which side has the largest (or smallest) mode number in annihilation operators acting on $\ket{\psi}$. The procedure stops when the products of annihilation operators on both sides coincide, which is what we wanted to prove. The equation $L_1\ket{\psi} = L_2 \ket{\psi}$ tells us in particular that the coefficients of the vectors $\ket{e_1}$ and $\ket{e_2}$ in the expansion of $\ket{\psi}$ are related. Since this argument applies to any pair of Fock vectors $\ket{e_1}$ and $\ket{e_2}$, all coefficients in $\ket{\psi}$ are related and \eqref{eq: constraints highest energy state} has a unique solution in each ($N,M$)-block. 


\section{Eigenstates within the top eigenspace of $H_{\min}$}\label{secHmintop}

In the previous section, we constructed the joint top eigenvector of $H$ and $\Hm$ in every ($N,M$)-block. We shall now explore the structure of the remaining eigenvectors of $H$ that reside in the top eigenspace of $\Hm$.

Consider the subspace of states in an $(N,M)$-block annihilated by all $B_{js\alpha}$ for $s \ge1$ and call it $V^{(1)}_{NM}$. From \eqref{H1} and \eqref{HHmH0H1}, it follows that their eigenvalue under $\Hm$ equals $M^2$, which saturates the upper bound for $\Hm$ (but not necessarily for $H$). Similarly to \eqref{eq: constraints highest energy state}, these states satisfy 
\beq\label{eq: constraints highest Hmin subspace} 
 a_{k+1} a_{j-k+1}|\phi^{(1)}_{N,M}\rangle= a_{l+1} a_{j-l+1}|\phi^{(1)}_{N,M}\rangle,
\eeq
for any $j,k$ and $l$.
Following the same line of thought as in the uniqueness proof at the end of the previous section, one can show that the dimension of this subspace is at most $\min(N,M)$. Indeed, the constraints \eqref{eq: constraints highest Hmin subspace} can be used to relate coefficients of Fock vectors that have the same number of modes $\ad_0$, while vectors with a different eigenvalue for $\ad_0a_0$ cannot be related using these constraints. This leaves $\min(N,M)$ coefficients in the Fock expansion unconstrained.

The idea is now to construct a basis for $V^{(1)}_{NM}$ and diagonalize $H$ in that $\Hm$ eigenspace. Since $H$ and $\Hm$ commute and any eigenstate outside of $V^{(1)}_{NM}$ has a lower eigenvalue for $\Hm$, this subspace is invariant under $H$ as well. (Note that the highest energy state derived in the previous section belongs to $V^{(1)}_{NM}$.) The key observation that allows us to construct a basis for $V^{(1)}_{NM}$ is the fact that the constraints \eqref{eq: constraints highest energy state} and \eqref{eq: constraints highest Hmin subspace} are related by the action of the shift operator $S$. Therefore, any shifted highest energy state \eqref{eq: highest energy state final} would satisfy \eqref{eq: constraints highest Hmin subspace}. A basis for $V^{(1)}_{NM}$ can thus be constructed by considering the states \eqref{eq: highest energy state final} in the blocks $(k,M-k)$, applying the shift operator, and then populating mode 0 by acting with $a_0^\dagger$ repeatedly to reach the target $(N,M)$-block:
\begin{equation}
\label{eq: top eigenspace Hmin}
|\phi^{(1,k)}_{N,M}\rangle \equiv \frac{a_0^{\dagger (N-k)}}{\sqrt{(N-k)!}}S|\psi^{\max}_{k,M-k} \rangle .
\end{equation}
Note that $k$ ranges from 1 to $\min(N,M)$, as expected by the above counting argument. Within each block, the states \eqref{eq: top eigenspace Hmin} are linearly independent as they contain a different number of modes $\ad_0$. In conclusion, the states \eqref{eq: top eigenspace Hmin} provide us with a complete basis for the top eigenspace of $\Hm$.

The next step is to compute the matrix elements of $H$ in the basis \eqref{eq: top eigenspace Hmin} and diagonalize the resulting matrix. Fortunately, since a quartic resonant Hamiltonian can at most create or destroy one zero mode, $H$ is tridiagonal in the basis \eqref{eq: top eigenspace Hmin}. Therefore, it is enough to focus on cases $k'=k$ and $k'=k+1$, in combination with the hermiticity of $H$. It will be useful to decompose $H$ according to the number of appearances of the zero mode. Let us first focus on the case $k'=k$. The only terms in $H$ that contribute in that case have an identical number of $\ad_0$ and $a_0$. These terms can be conveniently written as 
\begin{equation}
    H \to  \frac12\sum_{n+m=k+l}^{n,m,k,l\ge 0} a^\dagger_{n+1} a^\dagger_{m+1} a_{k+1} a_{l+1}+ 2\ad_0a_0\hat{N} -\frac32\left(\ad_0a_0\right)^2 -\frac12\ad_0a_0  \, ,
    \label{eq: diagonal contribution H}
\end{equation}
such that 
\begin{align}
 \bra{\phi^{(1,k)}_{N,M}} H \ket{\phi^{(1,k)}_{N,M}}&= \bra{\psi^{\max}_{k,M-k}} H \ket{\psi^{\max}_{k,M-k}} + 2(N-k)N-\frac32 \left( N-k \right)^2 - \frac12 (N-k)\nonumber \\
   &= \left( \frac{N(N-1)}{2}-M \right)+\left(N+M+1\right)k-2k^2.
   \label{eq: H top diagonal}
\end{align}
We now turn to the case $k'=k+1$, where the terms in $H$ need to have a single $a_0$ present, and no other creation-annihilation operator with index 0, to give a nonzero contribution:
\begin{align}
    \bra{\phi^{(1,k+1)}_{N,M}} H \ket{\phi^{(1,k)}_{N,M}}&=\frac{\bra{  \psi^{\max}_{k+1,M-k-1}}S^\dagger a_0^{ (N-k-1)} H a_0^{\dagger (N-k)}S\ket{\psi^{\max}_{k,M-k}} \nonumber }{\sqrt{(N-k-1)!(N-k)!}} \, ,\\
    &=\frac{\bra{  \psi^{\max}_{k+1,M-k-1}}S^\dagger a_0^{ (N-k)} \left(\sum_{n,m=1}^\infty  \ad_n \ad_m a_{n+m} \right) a_0^{\dagger (N-k)}S\ket{\psi^{\max}_{k,M-k}} \nonumber }{\sqrt{(N-k-1)!(N-k)!}} \, , \\
    &=\sqrt{N-k}\bra{  \psi^{\max}_{k+1,M-k-1}} K_+\ket{\psi^{\max}_{k,M-k}}   \, ,
    \label{eq: odddiag K+}
\end{align}
with 
\begin{equation}
    K_+ = \sum_{n,m=0}^\infty \ad_n \ad_m a_{n+m+1}  \, .
    \label{eq: def K+}
\end{equation}
Using the constraints \eqref{eq: constraints highest energy state} on the bra-state, the action of this operator $K_+$ on the highest-energy states can be simplified to
\begin{align}
 \bra{\phi^{(1,k+1)}_{N,M}} H \ket{\phi^{(1,k)}_{N,M}} 
   &= \sqrt{N-k}\bra{\psi^{\max}_{k+1,M-k-1}} \ad_0 \sum_{j=0}^\infty (j+1) a^\dagger_{j} a_{j+1}  \ket{\psi^{\max}_{k,M-k}} \, , \nonumber \\
   &=\sqrt{N-k}\bra{\psi^{\max}_{k+1,M-k-1}} \ad_0 Z^\dagger \ket{\psi^{\max}_{k,M-k}}\, .
   \label{eq:offdiag matrix elements intermediate}
\end{align}
It is simplest to consider the action of this product of operators on the bra-state.
Since $a_0$ commutes with the constraints \eqref{eq: constraints highest energy state}, this operator transports the highest energy state from the block $(N,M)$ to the highest energy state in block $(N-1,M)$. Moreover, $a_0$ and $Z$ commute so that the action of $a_0$ in \eqref{eq:offdiag matrix elements intermediate} can be straightforwardly computed using the expression \eqref{eq: highest energy state final}
\begin{equation}
     a_0 \ket{\psi^{\max}_{N,M}} = \sqrt{\frac{N(N-1)}{N+M-1}} \ket{ \psi^{\max}_{N-1,M}} \, .
     \label{eq:action a0}
\end{equation}
In addition, we established above that $Z$ maps the highest energy state \eqref{eq: highest energy state final} in block $(N,M)$ to the highest energy state in block $(N,M+1)$. Using \eqref{eq: highest energy state final}, one can compute the proportionality factor as
\begin{equation}
\label{eq:action Zd}
    Z \ket{ \psi^{\max}_{N,M}} = \sqrt{(M+1)(N+M)}\ket{ \psi^{\max}_{N,M+1}}  .
\end{equation}
Substituting \eqref{eq:action Zd} and \eqref{eq:action a0} in \eqref{eq:offdiag matrix elements intermediate}, the off-diagonal matrix elements become
\begin{equation}\label{eq: H top offdiag}
    \bra{\phi^{(1,k+1)}_{N,M}} H \ket{\phi^{(1,k)}_{N,M}} = \sqrt{(N-k)(M-k)(k+1)k} = \bra{\phi^{(1,k)}_{N,M}} H \ket{\phi^{(1,k+1)}_{N,M}}.
\end{equation}
This completes the computation of the matrix elements of $H$ in the top eigenbasis of $H_{\min}$. 

Tridiagonal matrices analogous to the ones we have arrived at, with nonzero matrix elements given by \eqref{eq: H top diagonal} and \eqref{eq: H top offdiag}, have been diagonalized in \cite{squareinteger}. Remarkably, their spectrum follows a square integer sequence. This property therefore also defines the eigenvalues of $H$ in the top $\Hm$ eigenspace. In fact, all the eigenvalues of $H$ that we will be able to bring under some degree of analytic control are parts of square integer sequences of this sort. We now repeat the argument of \cite{squareinteger}, adapted to our needs, for completeness.

The claim is that the eigenvalues of the tridiagonal matrix given by \eqref{eq: H top diagonal} and \eqref{eq: H top offdiag} are
\begin{equation}
    E^{(1,k)} = \left( \frac{N+M+2-2k}{2}\right)^2  +c^{(1)}
    \label{eq: H spectrum top Hmin} \quad \text{for} \quad  k=1,\dots,\min(N,M) ,
\end{equation}
with $c^{(1)} \equiv N(N-1)/2-M-(N-M)^2/4$ and the superscript refers to the top eigenspace of $\Hm$. In the following, we shall focus on deriving the square integer part of \eqref{eq: H spectrum top Hmin} and therefore subtract the constant part $c^{(1)}$ from the diagonal elements \eqref{eq: H top diagonal} in our treatment below.

Before delving into the proof, it is convenient to rewrite the matrix elements in terms of the dimensionality of the basis $n \equiv \min(N,M) $ and the parameter $a \equiv |N-M|$, instead of $N$ and $M$. The following relations hold:
\begin{equation}
\max(N,M)  = a + n \quad \text{and} \quad N+M = \min(N,M) + \max(N,M) = 2n+a.
\end{equation}
Note that either $N = n$ and $M = a+n$ or $M = n$ and $N = a+n$.
The diagonal elements \eqref{eq: H top diagonal}, with the constant $c^{(1)}$ subtracted, can then be written as 
\begin{equation}
    a_k = -2k^2+(2n+a+1)k +a^2/4\quad \text{for} \quad k = 1,\cdots,n \, ,
    \label{eq: ak's}
\end{equation}
while the off-diagonal terms become
\begin{equation}
    b_k= \sqrt{(n-k)(n+a-k)(k+1)k} \quad \text{for} \quad k = 1,\cdots,n-1.
    \label{eq: bk's}
\end{equation}
Together, they make up a matrix that we will refer to as $H_n$.
We now want to show that the eigenvalues of this tridiagonal matrix $H_n$ are \begin{equation}
    \tilde{E}_k = \left( \frac{a}{2}+k\right)^2  
    \label{eq: H spectrum top Hmin with C=0} \quad \text{for} \quad  k=1,\dots,n .
\end{equation}

The proof works by induction \cite{squareinteger}. The statement is trivially true at $n = 1$, since in that case $a_1 = \tilde{E}_1$. We aim to show that the statement is true at $n$ assuming it holds at $n-1$, keeping $a$ fixed. 
At level $n$ we consider the decomposition $H_n\equiv \left( n+ {a}/{2}\right)^2 I-C_n$  and notice that the matrix $C_n$ can be written as the product $C_n \equiv A_nA_n^T$ with $A_n$ a lower bidiagonal matrix. Its diagonal elements $h_k$ and lower subdiagonal elements $r_k$ are respectively given by
\begin{align}
    h_k = \sqrt{(n - k ) ( n  + a -  k)}&\quad \text{for} \quad k=1,\cdots, n,\\
    r_k  = -\sqrt{( k + 1) k} \quad \text{for} \quad &k=1,\cdots, n-1.
\end{align}
The key observation is that $D_n \equiv  \left( n+{a}/{2} \right)^2 I-A_n^T A_n$, where the matrix $A_n$ and $A_n^T$ have swapped their positions, decomposes as
\begin{equation}D_n = 
\begin{pmatrix}
    &   &   & 0 \\
    & H_{n-1} & & \vdots \\
    &   &   & 0 \\
    0 & \cdots & 0 &  \left(n+{a}/{2}\right)^2
\end{pmatrix}.
\label{eq: tensor decomposition}
\end{equation}
By use of the singular value decomposition of $A_n$, it is straightforward to show that $A_n^T A_n$ and $A_n A_n^T$ share the same nonzero eigenvalues. We therefore find that $H_{n-1}$ and $H_n$ display the same $n-1$ lowest eigenvalues (at fixed value of $a$) and the additional eigenvalue for $H_n$ is $\tilde{E}_n$. This completes the proof.
Note that the above construction with the swapped order of $A$ and $A^T$ in the partner matrices $C$ and $D$ is strongly reminiscent of
the factorization method in quantum mechanics \cite{factorization}, with its connections to supersymmetric quantum mechanics.

The eigenvectors of $H_n$ can be derived in an analogous manner. We first focus on the highest energy vector, with energy $\tilde{E}_n$. This corresponds to the null direction of $C_n$, which satisfies $A_n^T v=0$. This condition is solved by 
\begin{equation}
    v_{k+1} =- \frac{h_k}{r_k} v_k \quad \text{for} \quad k = 1,\cdots,n-1 \,,
    \label{eq: solution highest}
\end{equation}
where we used that $h_n = 0$.
This uniquely specifies the highest energy eigenvector of $H_n$. The normalized components of this vector in the basis \eqref{eq: top eigenspace Hmin} are
$$
    v^{(1,1)}_k = \sqrt{\frac{n! (a + n)!}{(2n + a -1)!}} \sqrt{\frac{1}{k} \binom{n-1}{k-1} \binom{n+a-1}{k-1} } = \sqrt{\frac{M! N!}{(N+M -1)!}} \sqrt{\frac{1}{k} \binom{N-1}{k-1} \binom{M-1}{k-1} } \, .
$$
The resulting energy eigenvector is precisely the highest energy state $\ket{{\psi}^{\max}_{N,M}}$ derived in the previous section.

In order to find the other eigenvectors, we use the product decomposition \eqref{eq: tensor decomposition}. We first focus on the $(n-1)$-dimensional space in which $H_{n-1}$ is defined. Following identical steps as above, we find the highest energy state $v^{(n-1)}$ of this matrix using \eqref{eq: solution highest} and replacing $n\rightarrow n-1$. This state, embedded into an $n$-dimensional space by appending a 0 to its column vector representation, is also an eigenstate of $D_n$. It is straightforward to show that, if $v^{(n-1)}$ is an eigenvector of $A_n^TA_n$, then $A_nv^{(n-1)}$ is an eigenvector of $A_nA_n^T$ with the same eigenvalue. The resulting vector $A_nv^{(n-1)}$ is an eigenvector of $H_n$ with eigenvalue $\tilde{E}_{n-1}$, as required. This process can be repeated for any of the eigenvalues $\tilde{E}_k$. 

The result of this section is complete analytic understanding of the eigenvalues and eigenvectors of $H$ in the top eigenspace of $H_{\min}$. (For later reference, we shall denote the normalized eigenvectors associated with the energies \eqref{eq: H spectrum top Hmin} as $\ket{\psi^{(1,k)}_{N,M}}$.) Under the shift operator, these joint eigenstates are mapped to eigenstates of $H_{\min}$ with distinct eigenvalues. These eigenvalues can be found using \eqref{eq:shift joint eigenvectors}. Applying \eqref{eq:shift joint eigenvectors} to the eigenstates with energies \eqref{eq: H spectrum top Hmin} considered in block $(N,M-N$) yields the following sequence of $H_{\min}$ eigenvalues in block $(N,M)$ 
\begin{equation}
   E_{\min} = \left(M-k+1\right)^2+\left(k-1\right)^2 \, ,
   \label{eq: regular energies}
\end{equation}
with $k = 1, \dots, \min(N,M-N)$. These $H_{\min}$ eigenvalues are preserved when transporting the shifted states to higher $N$ blocks using $a^\dagger_0$. For each $k$, one could try to construct a subspace in the block $(N,M)$ in which to diagonalize $H$ following the construction \eqref{eq: construction Hmin eigenspace}. Of course, we are not guaranteed to already know all the states in lower blocks that would map to the eigenvalues \eqref{eq: regular energies} under the shift $S$, and $H$ might therefore not leave the constructed subspaces invariant. Suppose we could complete the states $\ket{\psi^{(1,k)}_{N,M}}$ to a joint eigenbasis of $H$ and $\Hm$ in lower blocks and construct \eqref{eq: construction Hmin eigenspace}. Following similar steps as in the previous section, we would find that the Hamiltonian $H$ has nonzero matrix elements in two cases only. First, the matrix elements between two identical states can be computed using \eqref{eq: diagonal contribution H} and is generally nonzero. For a pair of distinct states with the same number of zero modes turned on in \eqref{eq: construction Hmin eigenspace}, the respective contribution vanishes by orthogonality of states in the spaces $V_{NM}$. Second, pairs of states \eqref{eq: construction Hmin eigenspace} that differ by one mode $\ad_0$ are related by $H$ if and only if the states from which they were derived using the shift operator have a nonzero overlap through the operator $K_+$, as in \eqref{eq: odddiag K+}. All other matrix elements vanish. 


\section{Ladder operators}\label{secladder}

We have observed numerically that the square integer structure displayed by the top eigenspace of $H_{\min}$ is replicated in other eigenspaces.
A key component in the diagonalization process of $H$ is the behavior of $K_+$ defined by (\ref{eq: def K+}) on joint eigenstates of $H$ and $\Hm$, as we have just discussed. In this section, we show that $K_+$ has in fact an even more important role in the structure of eigenstates and eigenvalues: it is a ladder operator for the combination $2H+\Hm$. 

Introducing the Hermitian conjugate of $K_-\equiv K_+^\dagger$, the following commutation relation holds:
\begin{equation}
    [2H+H_{\min},K_{\pm}]= \pm K_{\pm}.
    \label{eq: ladder operator K-}
\end{equation}
Hence, $K_+$ ($K_-$) raises (lowers) the eigenvalues of $2H+H_{\min}$ by one unit. 
Note that the part of this commutation relation quintic in creation-annihilation operators cancels out between the contributions from $H$ and $H_{\min}$.
The remaining cubic part is precisely the right-hand side of (\ref{eq: ladder operator K-}). This automatically implies that the classical quantities corresponding to $K_\pm$ 
are conservation laws for the dynamics induced by $2H+\Hm$, while their function as ladder operators in the quantum theory arises purely from quantum corrections to the classical Poisson brackets and the quantum correction in $H_{\min}$.

To verify \eqref{eq: ladder operator K-}, we will prove an equivalent statement:
\begin{equation}
    [H_{\min},\tilde{J}_3] = 0,\qquad \tilde{J}_3\equiv\sum_{n,m=1}^\infty \ad_{m+n} a_{m} a_{n}.
    \label{eq: shifted ladder commutation relation}
\end{equation}
This commutation relation follows straightforwardly by first observing that 
$$J_3\equiv\sum_{n,m=0}^\infty \ad_{m+n} a_{m} a_{n}$$ 
commutes with $H_{\min}$. From \eqref{oddJs}, $J_3$ is defined by the commutator of $a_0$ and $H$, both of which commute with $H_{\min}$. The difference between $\tilde{J}_3$ and $J_3$ involves $ Na_0$ and a cubic operator made of $a_0$ and $a_0^\dagger$, all of which commute with $\Hm$. This proves \eqref{eq: shifted ladder commutation relation}. To relate \eqref{eq: shifted ladder commutation relation} to \eqref{eq: ladder operator K-}, we notice that shifting the mode numbers up by one unit $a_n \rightarrow a_{n+1}$ (and the same for $\ad$) sends $K_-$ to $\tilde{J}_3$ and $2H+\Hm$ to $\Hm$, up to extra terms involving $N$ and $M$ that are described by
(\ref{eq:shift joint eigenvectors}). One can straightforwardly commute these extra terms with $K_-$, producing the nonzero right-hand-side in \eqref{eq: ladder operator K-}. 

This ladder operator has a complementary action to the operator $a^{\dagger}_0$ which acts as a ladder operator for $\Hm$ by copying its eigenvalues from block $(N,M)$ to $(N+1,M)$, as well as the shift operator $S$ whose action on the eigenvectors is determined by (\ref{eq:shift joint eigenvectors}). Of course, all the operators obtained by commuting either $K_{\pm}$ or $\ad_0$ with any of the conservation laws of the model trivially define additional (higher polynomial) ladder operators.


\section{Eigenstates within lower eigenspaces of $H_{\min}$}\label{secHmintop2}

Having understood the structure of the top eigenspace of $H_{\min}$, we would like to follow similar steps for the $H_{\min}$ eigenspace associated to the next eigenvalue in the sequence \eqref{eq: regular energies},
\begin{equation}
    E^{(N,M)}_{min}=(M-1)^2+1.
    \label{eq: energy second Hmin block}
\end{equation}
To this end, we consider the states
\begin{equation}
\label{eq: second eigenspace Hmin}
\ket{\phi^{(2,k)}_{N,M}} \equiv \frac{a_0^{\dagger (N-k-1)}}{\sqrt{(N\!-\!k\!-\!1)!}}S\ket{\psi^{(1,2)}_{k+1,M-k-1}}
\end{equation}
where $\ket{\psi^{(1,2)}_{N,M}}$ is the next-to-highest energy eigenvectors of $H$ (constructed as $A_nv^{(n-1)}$ in the previous section) in the top eigenspace of $\Hm$ in block $(N,M)$. The states \eqref{eq: second eigenspace Hmin} exist for $1\leq k \leq \min(N-1,M-3)$. Equation \eqref{eq: regular energies} shows that the states \eqref{eq: second eigenspace Hmin} all share the same $H_{\min}$ eigenvalue equal to \eqref{eq: energy second Hmin block} and we aim to diagonalize the Hamiltonian in this subspace. However, in order to perform this diagonalization, a complete basis of the eigenspace at energy \eqref{eq: energy second Hmin block} is needed. In contrast to section~\ref{secHmintop}, we do not have a counting argument showing that the states \eqref{eq: second eigenspace Hmin} form a complete basis for the eigenspace of $\Hm$ at value \eqref{eq: energy second Hmin block}. However, as we shall see below, the Hamiltonian $H$ indeed leaves the subspace spanned by the basis \eqref{eq: second eigenspace Hmin} invariant due to the simple action of $K_+$ on states $\ket{\psi^{(1,2)}_{N,M}}$, see \eqref{eq: action K+ on second highest}.

Evaluating $H$ in the basis \eqref{eq: second eigenspace Hmin} once more yields a tridiagonal matrix, with eigenvalues following the square integer sequence:
\begin{equation}
E^{(2,k)} = \left( \frac{N+M-2-2k}{2} \right)^2+\frac{N(N+1)}{2}-M- \frac{(M-N-2)^2}{4} ,
\label{eq: Hsz energies second subspace}
\end{equation}
with $k = 1,\cdots,\min(N-1,M-3)$. To show this, we construct the matrix elements of $H$ in \eqref{eq: second eigenspace Hmin}. The computation of the diagonal elements is identical to section~\ref{secHmintop}. It now takes as input the energy of the states $|\psi^{(1,2)}_{k+1,M-k-1} \rangle $,
\begin{equation}
E^{(1,2)}_k = (k-1) \left( M-1-\frac{k}{2}\right),
\end{equation}
so that 
\begin{equation}
  \bra{ \phi^{(2,k)}_{N,M}} H \ket{\phi^{(2,k)}_{N,M}} 
    =-2k^2+(N+M-3)k+ \frac{N(N+1)}{2}-M.
   \label{eq: H diagonal second}
\end{equation}
For the off-diagonal term with $k'=k+1$, we write
\begin{equation}
  \bra{ \phi^{(2,k+1)}_{N,M}} H \ket{\phi^{(2,k)}_{N,M}} 
   = \sqrt{N-k-1}\bra{\psi^{(1,2)}_{k+2,M-k-2}}K_+ \ket{\psi^{(1,2)}_{k+1,M-k-1}} =\bra{ \phi^{(2,k)}_{N,M}} H \ket{\phi^{(2,k+1)}_{N,M}}  \, ,
   \label{eq: offdiag second}
\end{equation}
The operator $K_+$ can be shown to have a simple action on the states $\ket{\psi^{(1,2)}_{n,m}}$:
\begin{equation}
    K_+ \ket{\psi^{(1,2)}_{N,M}} = \sqrt{N(N-1)(M-2)} \ket{\psi^{(1,2)}_{N+1,M-1}} \, .
    \label{eq: action K+ on second highest}
\end{equation}
The steps to show \eqref{eq: action K+ on second highest} are straightforward but tedious, and we summarize this derivation in Appendix~\ref{appK}. In light of the previous section, the map \eqref{eq: action K+ on second highest} is only possible by the subtle relation between the eigenvalue of the pair of states with respect to $2H+\Hm$, which have to differ by exactly one unit to be consistent with \eqref{eq: ladder operator K-}. 

We can then write \eqref{eq: offdiag second} as
\begin{equation}
    \bra{ \phi^{(2,k+1)}_{N,M}} H \ket{\phi^{(2,k)}_{N,M}} = \sqrt{(N-k-1)(M-k-3) ( k+1)k }.
    \label{eq: offdiag second final}
\end{equation}
with $k =1,\cdots, \min(M-3,N-1)-1$. 
The resulting matrix elements \eqref{eq: H diagonal second} and \eqref{eq: offdiag second final} have a form identical to \eqref{eq: H top diagonal} and \eqref{eq: H top offdiag}. This matrix can diagonalized in an analogous manner to  section~\ref{secHmintop}, now with $n = \min(N-1,M-3)$ and $a = |M-N-2|$. All subsequent steps are identical and one finds the square integer spectrum \eqref{eq: Hsz energies second subspace}, with eigenvectors, for the operator $H$ in the $\Hm$ eigenspace at level \eqref{eq: energy second Hmin block}. 

We can now shift the joint (subspace) eigenbasis $\ket{\psi^{(2,k)}_{N,M-N}}$ of $H$ and $\Hm$ using $S$ and find that the resulting $H_{\min}$ eigenvalues in block $(N,M)$ belong to the sequence
\begin{equation}
   E_{\min} = (M-k-1)^2+(k+1)^2,
   \label{eq: regular energies second}
\end{equation}
with $k = 1, \dots, \min(N-1,M-N-3)$. Note that there are generally at least two different types of states that are mapped to the $E^{(N,M)}_{\min} = (M-i)^2+i^2$ eigenspace (with $i>1$) under the shift: the states $\ket{\psi^{(1,1+i)}_{N,M-N}}
$ and the states $\ket{\psi^{(2,i-1)}_{N,M-N}}$. This means, using \eqref{eq:shift joint eigenvectors}, that these states share the same eigenvalue under $2H+\Hm$. This level is therefore degenerate, and $K_+$ will generically not be diagonal with respect to these two types of joint eigenvectors of $H$ and $\Hm$. This fact represents an obstacle for straightforwardly applying the above method for diagonalizing $H$ in further lower eigenspaces of $\Hm$. Indeed, the Hamiltonian $H$ in a basis composed of shifted joint eigenvectors will no longer be tridiagonal. Nonetheless, we have empirically found that the square integer sequences continue to emerge, suggesting that there exists a preferred basis for $K_+$ in which the $H$ eigenblocks acquire the tridiagonal structure of the sort encountered above.

We conclude this section with closed-form expressions for some of the numerically observed sequences of eigenvalues of $H$. At level $E_{\min} = \left( M - i\right)^2 +i^2$, with $i>0$ there are at least $i$ distinct square integer sequences parameterized by $n=0,\dots,i-1$ that follow the pattern
\begin{align}
   E^{i,n}_{k} = &\left( \frac{N+M-3i+n+1-2k}{2}\right)^2 + \frac{(N+n)(N+n+1)}{2} \nonumber \\ & \hspace{2.5cm} -M+ \frac{(1+n)(n-2i+2)}{2}-\left( \frac{M-N-n-i-1}{2} \right)^2 \, ,
\end{align}
for $k =1,\dots,\min(N-i+n,M-2i-1)$. This formula reproduces in particular the energies \eqref{eq: Hsz energies second subspace} when $i=1$. In addition, at level $E_{\min}=  (M-3-i)^2+(i+3)^2-2(i+2)$ for $i>0$ there are at least $i$ distinct square integer sequences parameterized by $n=0,\dots,i-1$ that are distributed according to the sequences:
\begin{align}
   E^{i,n}_{k} = &\left( \frac{N+M-3i+n-5-2k}{2}\right)^2 + \frac{(N+n+1)(N+n+2)}{2} \nonumber \\ & \hspace{2.5cm} -M+ \frac{(n+1)(n-2i-2)}{2}-\left( \frac{M-N-n-i-5}{2} \right)^2 \, ,
\end{align}
for $k =1,\dots,\min(N-i+n-1,M-2i-6)$.

\section{Relation to classical invariant manifolds}\label{secinvman}

The families of eigenstates derived above are reminiscent of the integrable sectors in the resonant systems studied in \cite{qperiod1,qperiod2}. These systems were previously shown to have a single (three-dimensional) invariant manifold \cite{BBCE,BEL,solvable} and the exactly solvable set of Hamiltonian eigenstates can be combined into coherent-like superpositions that reproduce the classical dynamics in the semiclassical limit \cite{qperiod1,qperiod2}. This connection between special subsets of Hamiltonian eigenstates and classical dynamically invariant manifolds can furthermore be effectively visualized using phase space techniques \cite{phase}. There is a considerable resemblance between this story outlined in  \cite{qperiod1,qperiod2} and the special families of eigenvectors of the GG Hamiltonian constructed above, hence one may expect the eigenstates of $H$ to include sectors that encode the classical dynamics within the invariant manifolds (\ref{invM}). 

The simplest correspondence of this sort is the relation between the top eigenstates of $H$ within each $(N,M)$-block and the ansatz \eqref{invM} with $R=1$. 
To see this, we examine the expectation values of the mode occupation number $\ad_n a_n$ within these states, which is an analog of the classical observable
$|\al_n|^2$.
For the highest energy state \eqref{eq: highest energy state final}, one can show that
\begin{equation}
     a_n |\psi^{\max}_{N,M} \rangle = \sqrt{N(N-1)}\,\,\sqrt{\frac{M!(N+M-n-2)!}{(M-n)!(N+M-1)!}}\,\,|\psi^{\max}_{N-1,M-n} \rangle \, .
\end{equation}
In the semiclassical limit, where $N,M \gg n $ are large while keeping their ratio $d^2 \equiv M/N$ fixed, one then gets 
\begin{equation}
    \langle \psi^{\max}_{N,M} | \ad_n a_n |\psi^{\max}_{N,M} \rangle = \frac{N}{1+d^2}\left(\frac{d^2}{1+d^2}\right)^n .
\end{equation}
This has exactly the same form as the observable $|\al_n|^2$ for $\al_n$ within the ansatz \eqref{invM} at $R=1$. The energy of the solutions in this lowest manifold, denoted by $\mathcal{M}(1)$ in \cite{GG}, likewise saturates the classical energy bound \cite{GG}.

We expect this connection with classical invariant manifolds to extend to more complicated families of eigenvectors, though we will not explore this in detail here.
To clarify our intuition, we recall that, according to \cite{GG}, in addition to the invariant manifolds \eqref{invM}, denoted there $\mathcal{M}(R)$, one has the invariant manifolds
\beq\label{invMtilde}
\al_n(t)=c(t)\,\de_{n,0}+\sum_{r=1}^R c_r(t)\,[p_r(t)]^n,
\eeq
denoted $\tilde{\mathcal{M}}(R)$.
There is an interlacing embedding of these manifolds into each other. The manifold $\mathcal{M}(R)$ is evidently a submanifold of $\tilde{\mathcal{M}}(R)$ obtained by setting $c=0$. On the other hand, the manifold $\tilde{\mathcal{M}}(R)$ is a submanifold of $\mathcal{M}(R+1)$ obtained by setting $p_{R+1}=0$, $c_{R+1}=c$. The action
of the classical shift operator (\ref{Scl}) moves configurations up this invariant manifold ladder. Namely, if we apply the shift to a configuration (\ref{invM}), the result still satisfies (\ref{invM}), with redefined $c_r$ and $p_r$, except for $n=0$, since now $\al_0=0$. In other words, we obtain a configuration that fits the ansatz (\ref{invM}), except for the shifted $\al_0$. But this is precisely the definition (\ref{invMtilde}), so the configuration is in $\tilde{\mathcal{M}}(R)$. Furthermore, since $\tilde{\mathcal{M}}(R)\in \mathcal{M}(R+1)$, subsequent action of the shift operator will push the configuration to $\tilde{\mathcal{M}}(R+1)$ and so on.

As explained above, the top eigenstates correspond to configurations in $\mathcal{M}(1)$ --- more precisely, one will have to take superpositions of the top eigenstates in different nearby blocks to obtain a state localized around a specific classical configuration $\al_n=cp^n$. We have furthermore been repeatedly obtaining other families 
of eigenvectors by acting on these top vectors (and their descendants) with the quantum shift operator (\ref{eq: relations p0 and S}). This strongly suggests that these lower families correspond to higher invariant manifolds described above. One would need, however, a more thorough and systematic understanding of the descendant eigenstate families to explore this correspondence comprehensively.


\section{Quantum Lax pair and conservation laws}\label{seclax}

Quantum Lax pairs \cite{ganoulis} are not a very common subject for discussion, since in general, they are both difficult to construct because of operator ordering problems and do not provide the same empowerment of the formalism and automatic construction of conservation laws as in the corresponding classical theory. They have found, however, some applications to quantum Calogero systems \cite{polychronakos}, see \cite{arutyunov,fairon}.

For the quantum GG Hamiltonian, we have been able to identify a curious Lax pair structure that we shall describe here. As in the classical case, introduce an auxiliary space $\vec{h}\equiv (h_0,h_1,\ldots)$ of number-valued vectors in the mode space. The Lax operators will be infinite matrices acting on these vectors, while the entries of these matrices will be Hilbert space operators made of $a$ and $a^\dagger$. The specific expressions are
\beq\label{defLMq}
(\mathcal{L}\vec{h})_n=\sum_{k,l=0}^\infty a_{l+k}^\dagger a_{k+n} h_l +nh_n,\qquad (\mathcal{M}\vec{h})_n=\sum_{l+m\ge n}^{l,m\ge 0}a_{l+m-n}^\dagger a_lh_m.
\eeq
Except for the second `quantum' term in $\mathcal{L}$, this replicates the structure of the classical Lax pair (\ref{defLMcl}).
Introducing the identity matrix $I$ in the space of $\vec{h}$, so that $I\vec{h}=\vec{h}$, the evolution of $\mathcal{L}$ is
\begin{equation}
    i \dot{\mathcal{L}} = [\mathcal{L},H I] \, ,
\end{equation}
and the Lax pair condition can be written simply as
\beq\label{qLax}
[HI+\mathcal{M},\mathcal{L}]=0.
\eeq

To prove by brute force that (\ref{qLax}) is valid, we can evaluate the commutators directly and then rewrite the result in terms of normal-ordered operators. The top-order piece, which is quartic in the creation-annihilation operators has to vanish by the classical computation in Appendix~\ref{appLax}, since all the terms are normal-ordered, no commutations of $a$ and $a^\dagger$ are necessary, and the classical and quantum computations will agree line-by-line. The remainder consists of two quadratic contributions from
\begin{align}
([\mathcal{M},\mathcal{L}]h)_n  = &\sum_{l+m\ge n}^{k,k',l,m\ge 0} a_{l+m-n}^\dagger a_l a_{k'+k}^\dagger a_{k+m} h_{k'} +\sum_{l+m\ge n}^{l,m\ge 0} m\,a_{l+m-n}^\dagger a_lh_m \nonumber \\
&-\sum_{k'+m\ge l}^{k,k',m,l\ge 0} a_{l+k}^\dagger a_{k+n} a_{k'+m-l}^\dagger a_{k'}h_m -n\sum_{l+m\ge n}^{l,m\ge 0}a_{l+m-n}^\dagger a_lh_m.
\label{eq: [M,L] quantum corrections cancel}
\end{align}
First, there are terms that emerge from bringing the quartic contributions to $[\mathcal{M},\mathcal{L}]$ into a normal-ordered form; second, there are terms arising from the quantum piece $nh_n$ in the $\mathcal{L}$-operator of (\ref{defLMq}). We must show that these two quadratic pieces cancel each other.
We consider the quartic contributions from the first and third terms in \eqref{eq: [M,L] quantum corrections cancel} in turn. The first term gives the following quadratic contribution:
\begin{align}
\sum_{l+m\ge n}^{k,k',l,m\ge 0} a_{l+m-n}^\dagger a_l a_{k'+k}^\dagger a_{k+m} h_{k'} &\to \sum_{l+m\ge n}^{k,k',l,m\ge 0} a_{l+m-n}^\dagger \delta_{l,k'+k} a_{k+m} h_{k'} = \sum_{k+k'+m\ge n}^{k,k',m\ge 0} a_{k+k'+m-n}^\dagger  a_{k+m} h_{k'} \nonumber \\
&= \sum_{\substack{k'+j\ge n\\ j\ge m}}^{j,k',m\ge 0} a_{k'+j-n}^\dagger  a_{j} h_{k'} = \sum_{k'+j\ge n}^{j,k'\ge 0}  j \, a_{k'+j-n}^\dagger  a_{j} h_{k'} .
\label{eq: contribution from ML}
\end{align}
First, we remove the sum over $l$ using the Kronecker $\delta$, then we introduce the index $j=k+m$ to swap the sum over $k$ for a sum over $j$. The label $m$ no longer appears in the summand, and therefore the respective sum is straightforwardly computed. We can perform similar steps for the third term in \eqref{eq: [M,L] quantum corrections cancel}:
\begin{align}
    \sum_{k'+m\ge l}^{k,k',m,l\ge 0} a_{l+k}^\dagger a_{k+n} a_{k'+m-l}^\dagger a_{k'}h_m  & \to \sum_{k'+m\ge l}^{k,k',m,l\ge 0} a_{l+k}^\dagger \delta_{k+n,k'+m-l}a_{k'}h_m = \sum_{\substack{k'+m\ge n \\ k'+m\ge n+k}}^{k,k',m\ge 0} a_{k'+m-n}^\dagger a_{k'}h_m \nonumber \\
    &= \sum_{k'+m\ge n}^{k',m\ge 0} (k'+m-n)a_{k'+m-n}^\dagger a_{k'}h_m ,
    \label{eq: contribution from LM}
\end{align}
where we performed the sum over $l$ and this eliminated all dependence on the label $k$ in the summand. Subtracting \eqref{eq: contribution from LM} from \eqref{eq: contribution from ML}, the remaining terms precisely cancel the second and fourth contributions in \eqref{eq: [M,L] quantum corrections cancel}. We have therefore explicitly shown that \eqref{qLax} holds at the quantum level, with the modified Lax pair \eqref{defLMq}.

It is usually impossible to directly construct quantum conservation laws from such quantum Lax pairs as $\Tr \mathcal{L}^n$, since the lack of commutativity of Hilbert space operators upsets the cyclic property of the trace \cite{ganoulis}. However, in our case, there is a trick that closely replicates the construction of classical conservation laws in (\ref{In}) originally discovered in \cite{GG}, which is different from the usual trace-based construction. Namely, we first introduce the vector $\vec{1}\equiv (1,0,0,0,\ldots)$ and notice the identity
\beq
\mathcal{L}\vec{1}=\mathcal{M}\vec{1}.
\eeq
Then we introduce the operators $(\vec{1},\mathcal{L}^n\vec{1})$ and observe that they commute with the Hamiltonian because 
\beq\label{qLaxcomm}
[H,(\vec{1},\mathcal{L}^n\vec{1})]=(\vec{1},[HI,\mathcal{L}^n]\vec{1})=(\vec{1},(\mathcal{L}^n\mathcal{M}-\mathcal{M}\mathcal{L}^n)\vec{1})=0.
\eeq

As the operators $(\vec{1},\mathcal{L}^n\vec{1})$ formally commute with the Hamiltonian, they provide an infinite family of conservation laws. There is unfortunately a complication with this construction, however, that precludes its immediate utilization for practical purposes. The operators obtained in this manner are not normal-ordered, and attempting to normal-order them, one gets contributions in the form of lower order polynomials in $a$ and $a^\dagger$ with infinite coefficients. It is natural to expect that these infinities are themselves lower-order conservation laws and may be systematically subtracted, yielding finite, well-defined conserved operators. Namely, one can truncate all sums at a finite large mode number and add a combination of lower-order conservation laws with divergent coefficients depending on this cutoff. With suitable finetuning, the total expression will have a well-defined limit as the cutoff is removed, providing a conservation law. We have verified using computer algebra that this can be done for the first few members of the tower. (We found the FORM software \cite{form} useful for this purpose, due to its built-in ability to effectively handle large polynomial expressions, possibly made of noncommuting variables.) We do not have, however, a systematic theory of such subtractions, and hence no way to employ the Lax pair (\ref{defLMq}) to construct conservation laws effectively.

In the course of our numerical experimentation, we have come across another construction of conservation laws that appears to work at all orders.
The formula is very simple and gives the conserved quantities as
\beq\label{J2n}
J_{2n} = [ J_{2n+1},a_0^\dagger],
\eeq
where the sequence $J_{2n+1}$ is defined recursively by (\ref{oddJs}). Note that $J_2$ is related to $N$, while $J_4$ is related to $H$. The higher $J$'s are independent polynomial conservation laws. To illustrate the sort of quantum corrections that are generated by \eqref{J2n}, we provide an explicit expression for $J_6$:
\begin{align}
   J_6 &=  2 \hspace{-0.3cm}\sum_{k, l, m,n,p=0}^\infty\hspace{-0.3cm}\ad_{k}  \ad_{l+m} \ad_{n+p} a_{k+l} a_{m+n} a_p +2 \hspace{-0.1cm}\sum_{k, l, m=0}^\infty (k+l+m) \, \ad_{k} \ad_{l+m} a_{k+l} a_{m} \\
   &\hspace{4cm}+2 \Hm + (2N+6)H + N^3 + N^2+2 M N + 2 M   .\nn
\end{align}
One can, of course, throw away the second line since it is expressed through the lower conservation laws. The sextic piece in the first line matches the structure of the classical conserved quantities (\ref{In}), but there is a nontrivial quartic correction.

We have not been able to prove analytically that (\ref{J2n}) are conserved, though we have run extensive checks that the corresponding matrices commute with the Hamiltonian within individual $(N,M)$-blocks. It is not difficult to understand, however, why the corresponding classical quantities are conserved as they are related to the classical tower (\ref{In}). This follows immediately from the expression for the classical quantities $J^{cl}_{2n+1}\equiv i\left\{J^{cl}_{2n-1},H_{cl}\right\}$ corresponding to the operators \eqref{oddJs}, with $J^{cl}_1\equiv \alpha_0$. In terms of the classical modes, they are 
\beq
 J^{cl}_{2n+1}= \hspace{-3mm}\sum_{i_1,\ldots, i_{2n}=0}^\infty \hspace{-3mm}\alpha_{i_1}\bar\alpha_{i_1+i_2} \alpha_{i_2+i_3}\cdots \bar\alpha_{i_{2n-1}+i_{2n}} \alpha_{i_{2n}}.
\label{classicaloddJs}
\eeq
Then, one straightforwardly finds the Poisson bracket
\beq
i\left\{J^{cl}_{2n+1} , \bar{\alpha}_0\right\}= \sum_{n=0}^{\infty} \frac{\partial J^{cl}_{2n+1}}{\partial \alpha_n} \frac{\partial \bar{\alpha}_0}{\partial \bar{\alpha}_n}  =  \sum_{i=0}^{n}  I_{n-i}I_i
\label{J2nclassical}
\eeq 
with $I_0 \equiv 1$, which is indeed conserved.
It is remarkable that applying the correspondence principle to the equivalent representation \eqref{J2nclassical} of the tower \eqref{I2n} to obtain \eqref{J2n} leads to quantum conservation laws without running into operator ordering issues.

Additionally, one expects an extra quantum tower related to the classical traces of powers of the Lax operator given by (\ref{Gn}). That tower starts with $M$ of (\ref{NMq}), and we also know its quartic member (\ref{Hmin}) that we discovered empirically in \cite{bound} and also provided a brute force proof that it is conserved in section~\ref{secHHmin}.
Additionally, we have recovered by trial-and-error as well as guided by the corresponding classical conservation law $G_3$ of  (\ref{Gn}) the following sextic conserved operator:
\begin{equation}
    \sum_{t,n,l,m,k,p,=0}^\infty \ad_{t+n}\ad_{l+m}\ad_{k+p}a_{n+l}a_{m+k}a_{p+t}
+ 6 \sum_{t,n,l,m=0}^\infty t \ad_{t+n}\ad_{l+m}a_{n+l}a_{m+t} + 
\sum_{n=0}^\infty n^3 \ad_{n}a_{n}.
\label{G3q}
\end{equation}
We do not know how to extend this tower to higher orders, nor is it clear what relation it bears to the quantum Lax pair (\ref{defLMq}).

Finally, perhaps the most remarkable property of all, which we once again only know from numerical observations, is that {\it all eigenvalues of the conservation laws are integers}, not only for $H$ and $H_{\min}$, but also for (\ref{J2n}) and (\ref{G3q}).


\section{Parallels with Calogero and Benjamin-Ono systems}

The presence of integer eigenvalues for the GG Hamiltonian and its partner conservation laws alludes to the structures typical of Calogero systems 
\cite{polychronakos, arutyunov, fairon}, the simplest of which describes $N$ particles on a line with $1/r^2$ two-body interactions:
\beq
H_{Calogero}=\sum_{i=1}^N \frac{p_i^2}2+\sum_{i<j}\frac{g}{(x_i-x_j)^2}.
\eeq
 Even beyond the mere observation of integer eigenvalues, the peculiar construction of conservation laws in
(\ref{In}) that does not utilize traces finds its parallels in considerations of Calogero particles \cite{fairon}.
Despite these intriguing similarities, the GG Hamiltonian is in the form of a field theory rather than a quantum-mechanical many-body system,
while straightforward second quantization of the Calogero system does not produce anything resembling the GG Hamiltonian \cite{calogero2q}.

More practical parallels can be seen between the structures revealed by our present exposition and the quantum Benjamin-Ono system \cite{quBO,Moll}. At the classical level,
similarity of integrability structures of the Benjamin-Ono and cubic \Sz{} equations has been revealed in \cite{GK,explicit} and has led to an explicit analytic solution
of the Benjamin-Ono equation analogous to (\ref{expl}) -- see \cite{CMNLS,halfwavemap} for further generalizations. The quantum Benjamin-Ono Hamiltonian defined in \cite{quBO} can be recast in a notation compatible with (\ref{HGG}) as
\beq\label{HqBO}
H_{qBO}=\sum_{m,n=1}^\infty \sqrt{\al^3 \,mn(m+n)}\left(\ad_m \ad_n a_{m+n}+\ad_{m+n}a_na_m\right)+\al(\al-1)\sum_{n=1}^\infty n^2\ad_n a_n,
\eeq
where $\al$ is a parameter. (Compared to \cite{quBO}, we have changed the normalization of the creation-annihilation operators in order to give them standard commutation relations.) This Hamiltonian conserves $M$ defined by (\ref{NMq}), but not the particle number $N$. It possesses a quantum Lax pair \cite{quBO} leading to a construction of conservation laws reminiscent of (\ref{qLaxcomm}), without encountering the divergences seen in the previous section for the GG Hamiltonian.

 A complete solution for the joint eigenstates of the Hamiltonian (\ref{HqBO}) and its hierarchy of conservation laws was given in \cite{quBO}.
The wavefunctions are expressed through Jack's polynomials in the variables $x_i$ defined by decomposing the classical variables $\al_n$ corresponding to the quantum operators $a_n$ as
\beq
\al_n=\frac1{\sqrt{n\al}}\left(x_1^n+x_2^n+x_3^n+\cdots\right),
\eeq
which is reminiscent of the formula (\ref{invM}) for the invariant manifolds of the GG Hamiltonian. The eigenstates of the Hamiltonian and other conserved operators
are indexed by partitions and the corresponding eigenvalues are expressed by elegant formulas in terms of part lengths \cite{quBO}.

Despite these tantalizing parallels between the structures observed in the GG Hamiltonian and the solution of the quantum Benjamin-Ono Hamiltonian in \cite{quBO},
we have not been able to make immediate use of these parallels. As already mentioned, the construction of conservation laws based on the quantum Lax pair for the GG Hamiltonian runs into divergences when attempting to normal-order the resulting operators. One can furthermore examine the numerically computed integer spectra within specific $(N,M)$-blocks and try to guess the corresponding formula in terms of part lengths. This does not appear possible, however, even for the very simple blocks with $N=2$ that contain only one nonzero eigenvalue \cite{quantres}. Intuitively, one should keep in mind the mode $a_0$ absent in the Benjamin-Ono Hamiltonian but present in the GG Hamiltonian. In the Fock states, the occupation number of this mode $\eta_0$ encodes the number of parts $N-\eta_0$ in the corresponding partition. Apparently, the eigenvalue formulas must be sensitive to the number of parts, and not only to the part lengths as in \cite{quBO}.


\section{Conclusions}

We have undertaken studies of integrability properties of the Hamiltonian (\ref{HGG}) and its partner Hamiltonian (\ref{Hmin}), relying on a mixture of analytic methods and numerical experimentation, and revealing rich and surprising algebraic structures. While we have not been able to solve the model in the sense of giving an explicit construction of all of its eigenfunctions and the corresponding eigenvalues, a number of prominent patterns have been identified. This includes the decompositions (\ref{boundHD}) and (\ref{boundHminD}) providing energy bounds, explicit construction of many families of eigenvectors in sections~\ref{sechighest}, \ref{secHmintop} and \ref{secHmintop2}, and alluding to their connections to classical invariant manifolds in section~\ref{secinvman}, the ladder operator (\ref{eq: def K+}), the quantum Lax pair (\ref{defLMq}) capable of generating higher conservation law, but requiring subtraction of divergent terms arising in the course of normal ordering, and empirically discovered conservation laws (\ref{J2n}) and (\ref{G3q}).

We hope that further studies of this system will supply the missing pieces of the puzzle, and similarities with the quantum Benjamin-Ono equation fully solved in \cite{quBO} are particularly promising in this regard. One ingredient that has been missing from our construction is the $r$-matrix theory, which has been systematically
used in some cases for generating the Hamiltonian eigenstates, cf.\ the considerations for quantum bosons with contact interactions in \cite{qub1,qub2}. There is also a distinction between the Lax pairs (\ref{defLMcl}) and (\ref{defLMq}) that our considerations revolve around, formulated as operators acting on an infinite-dimensional auxiliary space (this construction goes back to the original presentation due to G\'erard and Grellier in \cite{GG}), and finite-sized Lax matrices depending on a spectral parameter commonly seen in treatments of inverse scattering theory. It could be useful to find reformulations of our algebraic structures that would bridge this gap with other approaches.

The G\'erard-Grellier Hamiltonian forms a very special point in the space of theories (justifying the term `superintegrable' we have been employing to characterize it) and admits deformations preserving its Lax integrability \cite{Xu,cascade}. Once its quantum integrability has been brought fully under control, those deformations will emerge as natural further targets for study.


\section*{Acknowledgments}
We have benefited from discussions with Daniele Bielli, Lewis Cole, Patrick G\'erard, Sandrine Grellier, Enno Lenzmann, Alessandro Torrielli.
MDC is supported by a Leverhulme Early Career Fellowship and acknowledges partial support from STFC consolidated grant ST/X000664/1.
OE is supported by the  C2F program at Chulalongkorn University and by NSRF via grant number B41G680029.

\appendix

\section{The classical Lax pair}\label{appLax}

To verify that the classical Lax pair
\beq\label{defLMnrm}
(\mathcal{L}\vec{h})_n=\sum_{k,l=0}^\infty \al_{n+k} \ab_{k+l}h_l ,\qquad (\mathcal{M}\vec{h})_n=\sum_{m+p\ge n}^{m,p\ge 0}    \al_p\ab_{p+m-n} h_m.
\eeq
works properly, we must prove that
\beq\label{Laxtrgt}
i(\dot{\mathcal{L}}\vec{h})_n-(\mathcal{M}\mathcal{L}\vec{h})_n+(\mathcal{L}\mathcal{M}\vec{h})_n=0.
\eeq
Using the classical equation of motion
\beq
i\dot\al_n=\sum_{m=0}^\infty\sum_{p=0}^{n+m} \ab_m \al_p \al_{n+m-p}\, ,
\eeq
we have
\beq\label{Ldot}
i(\dot{\mathcal{L}}\vec{h})_n=-\sum_{k,l,m=0}^\infty \sum_{p=0}^{k+l+m} \ab_p \ab_{k+l+m-p}\al_m \al_{k+n}h_l   +\sum_{k,l,m=0}^\infty  \sum_{p=0}^{k+n+m} \ab_{l+k} \ab_m \al_p \al_{k+n+m-p} h_l.
\eeq
On the other hand,
\begin{align}
&(\mathcal{M}\mathcal{L}\vec{h})_n=\sum_{k,l=0}^\infty\sum_{m+p\ge n}^{m,p\ge 0} \ab_{m+p-n}   \ab_{l+k} \al_p \al_{k+m}  h_l ,\label{ML}\\
&(\mathcal{L}\mathcal{M}\vec{h})_n=\sum_{k,l=0}^\infty \sum_{m+p\ge l}^{m,p\ge 0} \ab_{l+k} \ab_{m+p-l}  \al_{k+n}  \al_p  h_m,\label{LM}
\end{align}
We first relabel the indices in (\ref{LM}) to bring the summand to the same form as in the first term of (\ref{Ldot}). Namely, we rename $m$ into $l$, $p$ into $m$, and $l$ into $p$, obtaining
\beq
(\mathcal{L}\mathcal{M}\vec{h})_n=\sum_{k,p=0}^\infty \sum_{l+m\ge p}^{l,m\ge 0} \ab_{p+k} \ab_{l+m-p}  \al_{k+n}  \al_m  h_l.
\eeq
We then introduce $\tilde p=p+k$ and drop tildes to obtain
\beq
(\mathcal{L}\mathcal{M}\vec{h})_n=\sum_{k=0}^\infty\sum_{p=k}^\infty \sum_{l+m\ge p-k}^{l,m\ge 0} \ab_{p} \ab_{k+l+m-p}  \al_{k+n}  \al_m  h_l.
\eeq
At this point, the summand is identical to the first term of (\ref{Ldot}), but the summation range can be expressed more effectively. The summation range is defined by the inequalities $p\ge k$, $l,m\ge 0$, $l+m\ge p-k$, which are more compactly written as $l,m\ge 0$, $k\le p\le k+l+m$, yielding
\beq
(\mathcal{L}\mathcal{M}\vec{h})_n=\sum_{k,l,m=0}^\infty\sum_{p=k}^{k+l+m} \ab_{p} \ab_{k+l+m-p}  \al_{k+n}  \al_m  h_l.\label{LMfnl}
\eeq
Then, we turn to (\ref{ML}) and attempt to redefine the indices to make the summand identical to the second term of (\ref{Ldot}). This is accomplished by introducing $\tilde m=m+p-n$, in terms of which the index range conditions $m,p\ge 0$, $m+p\ge n$ become $\tilde m\ge p-n$, $p\ge 0$, $\tilde m\ge 0$, or even 
better, $\tilde m\ge 0$, $0\le p\le \tilde m+n$. Dropping the tildes, we obtain
\beq
(\mathcal{M}\mathcal{L}\vec{h})_n=\sum_{k,l,m=0}^\infty\sum_{p=0}^{n+m} \ab_{m} \ab_{l+k} \al_p \al_{k+m+n-p}  h_l.\label{MLfnl}
\eeq
Finally, with (\ref{Ldot}), (\ref{LMfnl}) and (\ref{MLfnl}),
\begin{align}\label{Laxdiff}
(\dot{\mathcal{L}}\vec{h})_n-(\mathcal{M}\mathcal{L}\vec{h})_n+(\mathcal{L}\mathcal{M}\vec{h})_n=&-\sum_{k,l,m=0}^\infty \sum_{p=0}^{k-1} \ab_p \ab_{k+l+m-p}\al_m \al_{k+n}h_l  \\
 &+\sum_{k,l,m=0}^\infty  \sum_{p=n+m+1}^{k+n+m} \ab_{l+k} \ab_m \al_p \al_{k+n+m-p} h_l.\nonumber
\end{align}
In the second line, redefine $p$ to $n+m+k-p$, after that interchange $m$ and $p$, after that interchange the order of summations over $k$ and $m$ and define $\tilde k=k+p-m$, thereafter dropping the tildes and interchanging the order of summations of $k$ and $p$:
\begin{align*}
&\sum_{k,l,m=0}^\infty  \sum_{p=n+m+1}^{k+n+m} \ab_{l+k} \ab_m \al_p \al_{k+n+m-p} h_l =\sum_{k,l,m=0}^\infty  \sum_{p=0}^{k-1} \ab_{l+k} \ab_m \al_p \al_{k+n+m-p} h_l\\
&\hspace{1cm}=\sum_{k,l,p=0}^\infty  \sum_{m=0}^{k-1} \ab_{l+k} \ab_p \al_m \al_{k+n+p-m} h_l=\sum_{m,l,p=0}^\infty  \sum_{k=m+1}^{\infty} \ab_{l+k} \ab_p \al_m \al_{k+n+p-m} h_l\\
&\hspace{2cm}=\sum_{m,l,p=0}^\infty  \sum_{\tilde k=p+1}^{\infty} \ab_{l+\tilde k+m-p} \ab_p \al_m \al_{\tilde k+n} h_l=\sum_{k,l,m=0}^\infty  \sum_{p=0}^{k-1} \ab_{k+l+m-p} \ab_p \al_m \al_{k+n} h_l.
\end{align*}
Thus, the first and second lines of (\ref{Laxdiff}) are identical with a relative minus sign, and hence (\ref{Laxtrgt}) holds, ensuring the validity of the classical Lax pair.


\section{Some further details for the proof that $[H,H_{\min}]=0$}\label{apphhmin}

To complete the proof in section~\ref{secHHmin} that $[H,H_{\min}]=0$, we need to demonstrate that $A$ defined by (\ref{Adef}) satisfies $A_{kjj'}+A_{kj'j}+A_{j'jk}=0$.
To do so, we introduce the position space field $u(z)$ according to
\beq
a_n=\frac1{2\pi i}\oint \frac{dz}{z^{n+1}} u(z),\qquad u(z)\equiv\sum_{n=0}^\infty a_n z^n,
\eeq
so that
\beq\label{contour}
\begin{split}
&A_{kjj'}=\frac1{(2\pi i)^3} \oint \frac {dz}{z}\oint \frac {dv}{v}\oint \frac {dw}{w} A_{kjj'}(z,v,w),\\
&A(z,v,w)\equiv\sum_{l=0}^{j+k}\sum_{l'=0}^{j'+l} (\mathrm{min}_{j'+l,ll'}-\mathrm{min}_{j+k,lk})z^{l'}v^{j+k-l}w^{j'+l-l'}.
\end{split}
\eeq
To evaluate $A_{kjj'}(z,v,w)$, we need $\sum_{m=0}^n x^m=(1-x^{n+1})/(1-x)$, together with
\begin{align*}
&\sum_{m=0}^{k+l} \mathrm{min}_{k+l,lm} x^m\equiv\sum_{m=0}^{k+l} \min(k,l,m,k+l-m) x^m=\sum_{m=0}^{k+l} \sum_{s=1}^{\min(k,l,m,k+l-m)} x^m\\
&=\sum_{s=1}^{\min(k,l)}\sum_{m=s}^{k+l-s}  x^m=\sum_{s=1}^{\min(k,l)}x^s\sum_{m=0}^{k+l-2s}x^m=\sum_{s=1}^{\min(k,l)}\frac{x^s-x^{k+l-s+1}}{1-x}\\
&=\sum_{s=0}^{\min(k,l)-1}\frac{x^{s+1}-x^{k+l-s}}{1-x}
=\frac1{1-x}\left[x\frac{1-x^{\min(k,l)}}{1-x}-x^{k+l}\frac{1-x^{-\min(k,l)}}{1-1/x}\right]\\
&=\frac{x}{(1-x)^2}(1-x^{\min(k,l)}-x^{\max(k,l)}+x^{k+l})=\frac{x(1-x^{\min(k,l)})(1-x^{\max(k,l)})}{(1-x)^2}\\
&\hspace{3cm}=\frac{x(1-x^k)(1-x^l)}{(1-x)^2}.
\end{align*}
Then,
\begin{align}
&\sum_{l=0}^{j+k}\sum_{l'=0}^{j'+l} \mathrm{min}_{j'+l,ll'}z^{l'}v^{j+k-l}w^{j'+l-l'}=v^{j+k}w^{j'}\sum_{l=0}^{j+k}\frac{w^l}{v^l}\frac{z(1-z^{j'}/w^{j'})(1-z^l/w^l)}{w(1-z/w)^2}\nn\\
&=\frac{zv^{j+k}w^{j'+1}}{(w-z)^2}(1-z^{j'}/w^{j'})\sum_{l=0}^{j+k}\left(\frac{w^l}{v^l}-\frac{z^l}{v^l}\right)\nn\\
&=\frac{zwv^{j+k}}{(w-z)^2}(w^{j'}-z^{j'})\left(\frac{1-(w/v)^{j+k+1}}{1-w/v}-\frac{1-(z/v)^{j+k+1}}{1-z/v}\right)\nn\\
&=\frac{zw}{(w-z)^2}(w^{j'}-z^{j'})\left(\frac{v^{j+k+1}-w^{j+k+1}}{v-w}-\frac{v^{j+k+1}-z^{j+k+1}}{v-z}\right)\nn\\
&=\frac{v^{j+k+1}zw(w^{j'}-z^{j'})}{(w-z)(v-w)(v-z)}-\frac{zw^{j+k+2}(w^{j'}-z^{j'})}{(w-z)^2(v-w)}+\frac{z^{j+k+2}w(w^{j'}-z^{j'})}{(w-z)^2(v-z)}.
\label{A1}
\end{align}
Similarly,
\begin{align}
&\sum_{l=0}^{j+k}\sum_{l'=0}^{j'+l} \mathrm{min}_{j+k,lk}z^{l'}v^{j+k-l}w^{j'+l-l'}=v^{j+k}w^{j'}\sum_{l=0}^{j+k}
\mathrm{min}_{j+k,lk}\frac{w^l}{v^l}\frac{1-(z/w)^{j'+l+1}}{1-z/w}\nn\\
&=\frac{v^{j+k}w^{j'+1}}{w-z}\sum_{l=0}^{j+k}
\mathrm{min}_{j+k,lk}\frac{w^l}{v^l}-
\frac{v^{j+k}z^{j'+1}}{w-z}\sum_{l=0}^{j+k}
\mathrm{min}_{j+k,lk}\frac{z^l}{v^l}\nn\\
&=\frac{vw^{j'+2}(v^j-w^j)(v^k-w^k)}{(w-z)(v-w)^2}-
\frac{vz^{j'+2}(v^j-z^j)(v^k-z^k)}{(w-z)(v-z)^2}\label{A2}
\end{align}
By subtracting (\ref{A2}) from (\ref{A1}), we obtain $A_{kjj'}(z,v,w)$. Since the $(z,v,w)$-integration in (\ref{contour}) is fully symmetric under all permutations of $z$, $v$ and $w$, while $A_{kjj'}(z,v,w)$ is only symmetric with respect to interchanging $z$ and $w$ it is wise to equivalently consider the symmetrized combination
\begin{align*}
&A^{\mathrm{symm}}_{kjj'}(z,v,w)\equiv \frac13\left[A_{kjj'}(z,v,w)+A_{kjj'}(v,z,w)+A_{kjj'}(z,w,v)\right]\\
&=\frac{zvw}{3(w-v)(w-z)(v-z)} \left[P_{kjj'}(v,z) + P_{kjj'}(z,w) +P_{kjj'}(w,v) \right]\\
&-\frac{wv^2}{3(v-z)(v-w)^2}\left[P_{kjj'}(v,w) + P_{kjj'}(w,v)\right]-\frac{zw^2}{3(w-z)^2(w-v)}
\left[P_{kjj'}(z,w) + P_{kjj'}(w,z)\right]\\
&\hspace{3cm}+ \frac{vz^2}{3(w-z)(v-z)^2}\left[ P_{kjj'}(z,v) + P_{kjj'}(v,z) \right],
\end{align*}
with $P_{abc}(x,y)\equiv 2x^{a+b}y^c-x^{b+c}y^a-x^{a+c}y^b$. 
But since $P_{abc}+P_{acb}+P_{cba}=0$,
$A^{\mathrm{symm}}_{kjj'}(z,v,w)+A^{\mathrm{symm}}_{kj'j}(z,v,w)+A^{\mathrm{symm}}_{j'jk}(z,v,w)=0$, 
and hence by (\ref{contour}), 
$A_{kjj'}+A_{kj'j}+A_{j'jk}=0$. 
This proves that the last term in the square brackets of (\ref{H0H1quart}) gives a vanishing contribution, while all the other terms have been proven earlier to give vanishing contributions as well, hence $[H_0,H_1]=0$, and therefore $[H,H_{\min}]=0$.


\section{Action of $K_\pm$}\label{appK}
In this appendix we prove that $K_+$ \eqref{eq: def K+} maps between next-to-highest eigenvectors in the top $\Hm$ eigenbasis, i.e.,
\begin{equation}
    K_+ \ket{\psi^{(1,2)}_{N,M}} = \sqrt{N(N-1)(M-2)} \ket{\psi^{(1,2)}_{N+1,M-1}} \, ,
    \label{eq: action K+ on second highest app}
\end{equation}
with 
\begin{equation}
   \ket{\psi^{(1,2)}_{N,M}} = \sum_{k=1}^{\min(N,M)} v^{(1,2)}_{k,{\scriptscriptstyle{NM}}}\frac{a_0^{\dagger (N-k)}}{\sqrt{(N-k)!}}S|\psi^{\max}_{k,M-k} \rangle \, .
\end{equation}
We start by noting that $K_+$ commutes with $\ad_0$ and transporting $K_+$ through the shift operator gives
\begin{equation}
    K_+S = SK_+^{(2)} + 2a^\dagger_0S\tilde{Z}^\dagger+a^{\dagger2}_0Sa_0
    \label{eq: K+ through S}
\end{equation}
with 
\begin{equation}
  K_+^{(2)} \equiv \sum_{n,m=0}^\infty \ad_n\ad_m a_{n+m+2} \, , \qquad \tilde{Z}^\dagger \equiv \sum_{n=0}^\infty \ad_n a_{n+1} \, .
\end{equation}
The operator $K_+$ is hence found to be tridiagonal in the basis $|\phi^{(1,k)}_{N,M}\rangle$ given by \eqref{eq: top eigenspace Hmin}.\footnote{This is a slight abuse of language as $K_+$ is generally a rectangular matrix in these bases.} Let us compute each contribution separately. First, it is straightforward to show by direct computation that 
\begin{equation}
    K^{(2)}_+ \ket{\psi^{\max}_{N,M}} = \sqrt{\frac{M(M-1)N(N+1)}{N+M-1}} \ket{\psi^{\max}_{N+1,M-2}}
    \label{eq app:action K+2}
\end{equation}
using the explicit expression for the highest energy state \eqref{eq: highest energy state final} and the following commutation relations
\begin{equation}
    [K^{(2)}_+,Z] = 2K_+ \, , \qquad [K_+,Z] = J_3^\dagger \,  \qquad  \text{and} \qquad [J_3^\dagger,Z] = 0 \, .
\end{equation}
The second contribution in \eqref{eq: K+ through S} can be found using 
\begin{equation}
    [\tilde{Z}^\dagger,Z^m]=mNZ^{m-1} \,,
\end{equation}
which leads to
\begin{equation}
    \tilde{Z}^\dagger \ket{\psi^{\max}_{N,M}} = N \sqrt{\frac{ M}{N+M-1}}\ket{\psi^{\max}_{N,M-1}}\, .
    \label{eq app:action Ztilde}
\end{equation}
Finally, the third contribution to \eqref{eq: K+ through S} is easily found using \eqref{eq:action a0}, which we copy here for ease
\begin{equation}
     a_0 \ket{\psi^{\max}_{N,M}} = \sqrt{\frac{N(N-1)}{N+M-1}} \ket{ \psi^{\max}_{N-1,M}} \, .
     \label{eq app:action a0}
\end{equation}
Combining \eqref{eq: K+ through S}, \eqref{eq app:action K+2}, \eqref{eq app:action Ztilde} and \eqref{eq app:action a0} one finds 
\begin{align}
    K_+ \ket{\psi^{(1,2)}_{N,M}} = &\sum_{k=1}^{\min(N,M-2)} v^{(1,2)}_{k,{\scriptscriptstyle{NM}}} \sqrt{\frac{(M-k)(M-k-1)k(k+1)}{M-1}}  \frac{a_0^{\dagger (N-k)}}{\sqrt{(N-k)!}} S\ket{\psi^{\max}_{k+1,M-k-2}} \, ,\nonumber \\
    + &\sum_{k=1}^{\min(N,M-1)} 2v^{(1,2)}_{k,{\scriptscriptstyle{NM}}} k \sqrt{\frac{M-k}{ M-1}}  \frac{a_0^{\dagger (N-k+1)}}{\sqrt{(N-k)!}}S\ket{\psi^{\max}_{k,M-k-1}} \, , \nonumber \\
    + &\sum_{k=2}^{\min(N,M)} v^{(1,2)}_{k,{\scriptscriptstyle{NM}}} \sqrt{\frac{k(k-1)}{M-1}}  \frac{a_0^{\dagger (N-k+2)}}{\sqrt{(N-k)!}}S\ket{\psi^{\max}_{k-1,M-k}} \, .
    \label{eq: K+ intermediate}
\end{align}
The coefficients $v^{(1,2)}_{k,{\scriptscriptstyle{NM}}}$ can be computed following the recursion steps detailed at the end of section \ref{secHmintop}: 
\begin{align}
    v^{(1,2)}_{k,{\scriptscriptstyle{NM}}} &= \frac{(2n+a-1)k-n(n+a)}{\sqrt{k(2n+a-1)(2n+a-3)}} \sqrt{\frac{(n-2)! (n+a-2)!}{(2n + a -4)!}} \sqrt{\binom{n-1}{k-1} \binom{n+a-1}{k-1} } \nonumber \\
    &= \frac{(N+M-1)k-NM}{\sqrt{k(N+M-1)(N+M-3)}} \sqrt{\frac{(N-2)! (M-2)!}{(N+M -4)!}} \sqrt{\binom{N-1}{k-1} \binom{M-1}{k-1} }  \, .
    \label{eq: coefficients second highest}
\end{align}
Substituting \eqref{eq: coefficients second highest} in \eqref{eq: K+ intermediate} and rearranging terms leads to \eqref{eq: action K+ on second highest app}.

\end{document}